\documentclass[aps,prd,onecolumn,superscriptaddress]{revtex4-2}
\usepackage{amsmath,amssymb,bm,physics,graphicx,mathtools,cancel,color}
\usepackage[T1]{fontenc}
\usepackage{lmodern}
\usepackage{booktabs}
\usepackage{hyperref}
\newcommand{\M}{{\mathsf{M}}}
\hypersetup{
  colorlinks=true, linkcolor=blue, citecolor=blue, urlcolor=blue,
  pdftitle={The wormhole phase of coupled double-scaled SYK caps long-chord order parameters},
  pdfauthor={Hao Dai, Yu-Ge Chen, Li-Guo Qin, Li-Jun Tian}
}

\begin{document}

\title{The wormhole phase of coupled double-scaled SYK\\
caps long-chord order parameters}

\author{Hao Dai}
\email{Isaac.H.Dai@outlook.com}
\affiliation{Department of Physics, Shanghai University, Shanghai 200444, China}
\author{Yu-Ge Chen}
\email{chenyuge@shu.edu.cn}
\affiliation{Institute for Quantum Science and Technology, Shanghai University,
Shanghai 200444, China}
\author{Li-Guo Qin}
\email{lgqin@foxmail.com}
\affiliation{School of Mathematics, Physics and Statistics, Shanghai University
of Engineering Science, Shanghai 201620, China}
\author{Li-Jun Tian}
\email{tianlijun@shu.edu.cn}
\thanks{Corresponding author.}
\affiliation{Department of Physics, Shanghai University, Shanghai 200444, China}

\begin{abstract}
We couple two Berkooz--Brukner--Jia--Mamroud (BBJM) models---double-scaled SYK
with an integrable commuting-SYK deformation---by a Maldacena--Qi bilinear, and
extend the cylinder-phase chord rules to the second chord species. In the
cylinder every chord carries a species-blind traveling penalty
$e^{-\mu L}$, with $L$ the chord length,
while crossings keep the BBJM weights. Because the penalty depends only on the
time separation the bilocal saddle collapses to one dimension, and the cylinder phase is, at the
level of the chord rules and the saddle, a BBJM chord model in a linear
confining potential of slope $\mu=p\eta$, with $\eta$ the coefficient of the
microscopic bilinear. The potential caps long-chord order parameters. For the
quasi-integrable phase the order parameter saturates at
$\lvert g_z(\beta/2)\rvert\simeq2\kappa^2/\mu^2$, against
$\kappa^2\beta^2/4$ in the disk; more generally any saddle whose order
parameter must grow like $\beta^2$ over the whole circle is unavailable. The response of the bulk geodesic length is therefore suppressed by a factor $20$--$2\times10^4$ relative to the disk phase, and for $d=2$ no flavor-broken saddle
survives above $\mu^{*}\simeq0.31$ for $\kappa=0.6$--$0.85$. The wormhole is
thus robust in the interior of its own phase, consistent with the reported
insensitivity of the teleportation signal to the destruction of chaos; the
reason is the geometry of the phase, not the sector probed. The phase
boundary itself follows from an action difference; at leading order the
Hawking--Page temperature is universal, $\eta\beta_{\rm HP}=2\,\mathrm{arccosh}(2)$,
with the deformation entering only at $\mathcal{O}(\lvert D\rvert/p^2)$.
\end{abstract}

\maketitle

\section{Introduction}

The Sachdev-Ye-Kitaev (SYK) model
\cite{Sachdev:1992fk,Kitaev2015,Maldacena:2016hyu,Chowdhury:2021qpy} is
solvable in its double-scaled limit (DSSYK) through chord diagrams, and it
provides one of the few settings in which a transition between chaotic and
integrable dynamics can be followed analytically
\cite{Berkooz:2018jqr,Berkooz:2018qkz,Lin:2022rbf}. The Berkooz--Brukner--Jia--Mamroud
(BBJM) model \cite{Berkooz:2024evs,Berkooz:2024ofm} deforms DSSYK by an integrable commuting-SYK Hamiltonian \cite{Gao:2023gta} acting
on the same fermions, and exhibits a first-order transition between a chaotic
and a quasi-integrable phase at $T=T_c(\kappa)$. The dynamics across that
transition---chord number, Krylov complexity and operator size---has been
characterized in Ref.~\cite{Aguilar-Gutierrez:2026jjv}; what we add here is
the coupling of two such dots, and the effect of the resulting wormhole on
the same chord-sector response.

Separately, the Maldacena--Qi (MQ) model \cite{Maldacena:2018lmt} couples two
SYK dots by a bilinear $i\eta\sum_i\psi_i^L\psi_i^R$. In the
double-scaled limit its phase diagram contains three saddles, holographically
a pair of Euclidean black-hole disks, a connected $\mathrm{AdS}_2$ cylinder (the
traversable wormhole) and a cylinder with a baby universe; the chord rules for
all three were given recently in Ref.~\cite{Sontag:2026}. Its chord
description follows Ref.~\cite{Lin:2022rbf}, and its relation to
two-dimensional dilaton gravity is discussed in Refs.~\cite{Saad:2018bqo,Blommaert:2024ydx}.

Deformations of the chord model along renormalization-group directions have
also been studied \cite{Anninos:2022qgy}, as have its spectral properties
\cite{Polchinski:2016xgd}.

The two deformations probe different things. BBJM changes the \emph{spectrum}
of each dot, interpolating between a chaotic and an integrable Hamiltonian at
fixed degrees of freedom. The MQ coupling changes the \emph{topology} of the
bulk: it competes two disconnected disks against a connected geometry. It is
therefore not obvious what happens when both are present. On the one hand the
wormhole requires a coherent two-sided coupling and one might expect
integrability, with its large set of conserved charges, to destroy it. On the
other hand the wormhole is a low-temperature phase, and the quasi-integrable
phase is also a low-temperature phase, so the two could reinforce each other.

The question we address has been raised before in other forms. Garc\'ia-Garc\'ia et al.\
studied the chaos transition of a model dual to an eternal traversable
wormhole, in the pure Maldacena--Qi setting without an integrable sector
\cite{Nosaka:2019}. More directly, Dubey has recently destroyed the chaos of
the two coupled dots by random coupling deletion and monitored the
teleportation signal across the chaos-to-integrable transition, finding that
the ensemble-averaged peak height is largely unaffected
\cite{Dubey:2026}. He also offers an explanation: the signal measures
information transfer, which is controlled by $H_{\rm int}$, while
sparsification modifies only $H_L$ and $H_R$, so the signal probes a sector
that the deformation leaves invariant. That explanation is structural---it
refers to which sector the observable sees---and it predicts a dependence on
the form of the coupling, a non-diagonal $H_{\rm int}$ being expected to break
the decoupling; the test is left open in Ref.~\cite{Dubey:2026}.

Our question is at a different layer. We do not compute a transmission signal;
we ask how the bulk geometry responds. In the chord model the answer is that
the cylinder phase imposes a linear confining potential, so the deformation is
not merely confined to another sector but is actively suppressed by the
geometry of the phase itself. That yields a cap $2\kappa^2/\mu^2$ on the order
parameter, which does not follow from a sector-counting argument and which a
finite-$N$ study could test.

In this paper we study the model in which each dot is a BBJM model and the two
dots are coupled by an MQ bilinear. The logic runs along a single chain. The
microscopic model fixes a chord algebra, whose two-species form we write
explicitly in Appendix~\ref{app:penalty}, Eq.~\eqref{eq:chordalg}. Its crossing rule is the two-species rule of
Refs.~\cite{Berkooz:2024evs,Berkooz:2024ofm}; the weights \eqref{eq:weights}
follow from the index-set overlaps counted in Appendix~\ref{app:penalty}. What
has to be established here is that the cylinder penalty is species-blind, that
the inner product is Eq.~\eqref{eq:chordweight}, and that the Hamiltonian is
self-adjoint. The
double-scaled limit of that algebra produces the collective-field action \eqref{eq:S}; the
traveling penalty of the cylinder enters that action as a chemical
potential for the chord number; and solving the resulting saddle equation
gives the cap, the geodesic response and the phase diagram that follow. We
work in the $\lambda\to0$ semiclassical limit of the chord path integral,
where the saddle point is controlled by the $1/\lambda$ in the exponent. Most of the analysis is at
one integrable flavor per dot; the two- and three-flavor cases are treated
explicitly in Sec.~\ref{sec:flavors}. Fig.~\ref{fig:model} summarizes this chain: the model, the chord rules it implies, and the mechanism by which the cylinder caps the order parameter. Our main results are:

\begin{enumerate}
  \item The cylinder phase of the coupled model is, at the level of the chord
  rules and the semiclassical saddle, a BBJM chord
  model placed in a linear confining potential of slope $\mu=p\eta$
  (Sec.~\ref{sec:cylinder}). The traveling penalty of
  Ref.~\cite{Sontag:2026} is a function of $\tau_1-\tau_2$ only, so
  translation invariance survives and the bilocal saddle still depends on
  $\tau=\tau_1-\tau_2$. The BBJM equations are unchanged; only their source is
  multiplied by $e^{-\mu\tau}$.
  \item The quasi-integrable order parameter is \emph{capped} in the cylinder:
  $|g_z(\beta/2)|\to(2\kappa^2/\mu^2)[1-\frac32\nu^2/\mu^2+\cdots]$ as
  $\mu\gg\nu,\kappa$, against
  $\kappa^2\beta^2/4$ in the disk phase (Sec.~\ref{sec:cap}). The wormhole
  therefore cannot host the disk-like quasi-integrable saddle.
  \item In spite of this the wormhole is \emph{not} destroyed at the
  chaos--integrability point, provided the Maldacena--Qi coupling is not too
  weak: comparing the two-disk and cylinder free energies, the boundary
  satisfies $\kappa_w\ge\kappa_c$ at every point where
  both quantities could be computed, with a comfortable margin for
  $\eta\gtrsim0.03$, once $\kappa_c$ is evaluated at the
  temperature that the corresponding $\lvert\tilde\Delta\rvert$ implies (at
  $\eta=0.02$ the two are comparable and the survival is marginal); the three
  phases meet at a triple point (Sec.~\ref{sec:phase}).
  \item The wormhole is almost transparent to the deformation in a direct
  geometric probe. In the disk phase the chord-number response $R=\delta(-g)$ grows by $\kappa^2\beta^2/4$, i.e.\ the geodesic length by $\frac{\lambda}{2}R$; in the cylinder $R$ is suppressed by factors of $20$--$2\times10^4$, so the thermal correlator is destroyed by integrability in the
  disk phase and essentially unaffected inside the wormhole
  (Sec.~\ref{sec:length}).
  \item The Hawking--Page temperature is universal at leading order,
  $\eta\beta_{\rm HP}=2\mathrm{arccosh}(2)$, independently of $\kappa$; the deformation
  enters only through the $\mathcal{O}(\lvert D\rvert/p^2)$ correction in
  Eq.~\eqref{eq:hpexact} (Sec.~\ref{sec:temperature}).
\end{enumerate}

Appendix~\ref{app:penalty} verifies that the integrable chords carry the same
traveling penalty as the chaotic ones, which is what makes the cylinder
description used in Sec.~\ref{sec:cylinder} consistent.

\section{The model and its chord rules}
\label{sec:model}

Each dot is a BBJM model \cite{Berkooz:2024evs},
\begin{equation}
  \hat H^{(s)} = \nu\,\hat H^{(s)}_{\rm DSSYK} + \kappa\,\hat H^{(s)}_{\rm int},
  \qquad \nu^2+\kappa^2=1 ,
  \label{eq:bbjm}
\end{equation}
with $\hat H^{(s)}_{\rm int}$ the commuting-SYK Hamiltonian
\cite{Gao:2023gta}. The two dots are coupled by a Maldacena--Qi bilinear,
\begin{equation}
  \hat H = \hat H^{(L)} + \hat H^{(R)}
  + i\frac{\mu}{p}\sum_{i=1}^{N}\psi_i^{(L)}\psi_i^{(R)} .
  \label{eq:H}
\end{equation}
We write $\eta\equiv\mu/p$ for the coefficient of this bilinear, so that the
single-particle energies it induces are $\pm\eta/2$. The two differ by a factor $p$ and
both are used below, so we keep them apart from the start: $\eta$ is the parameter that carries the topological term, while
$\mu=p\eta$ is the exponent of the traveling penalty and hence the
parameter that appears in every saddle equation and every cap. In the saddle
analysis of Secs.~\ref{sec:cylinder}--\ref{sec:length} the parameter that
matters is $\mu$; at $\mu=\mathcal{O}(1)$, which is the regime in which the
cap competes with the disk, $\eta=\mathcal{O}(1/p)$. The topological term,
by contrast, can be evaluated at any $\eta$, and
Secs.~\ref{sec:phase}--\ref{sec:temperature} use it as the free parameter it
is. Sec.~\ref{sec:cylinder} states the reason for the $p$-th power.
The symbol $\nu$ is reserved for the BBJM weight. We work with the
chord-level coupling $\kappa$ of the BBJM phase diagram, not with the
microscopic variance of the commuting-SYK couplings; the relation is
Eq.~\eqref{eq:kappaB}.

In the double-scaling limit $N,p\to\infty$ with $\lambda=2p^2/N$ fixed, the
ensemble-averaged moments are sums over chord diagrams with, on each dot, two species of chords: chaotic ($n$) and integrable ($z$)
\cite{Berkooz:2024evs,Berkooz:2024ofm}, with crossing weights
\begin{equation}
  q_{nn}=q_{nz}=q=e^{-\lambda},\qquad q_{zz}=1 ,
  \label{eq:weights}
\end{equation}
i.e.\ the intersection matrix $\M_{nn}=\M_{nz}=1$, $\M_{zz}=0$.

\begin{figure}[t]
  \centering
  \includegraphics[width=\textwidth]{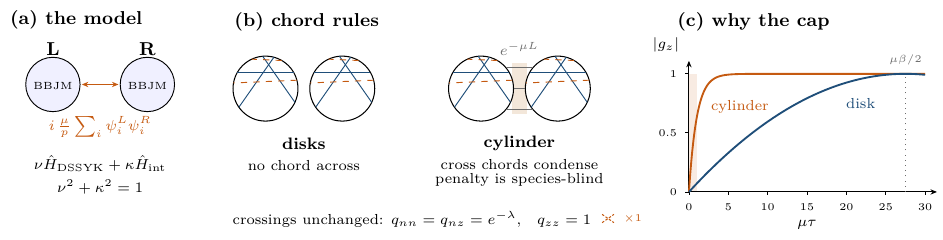}
  \caption{The model, its chord rules, and the origin of the cap. (a) Each dot is a
  BBJM model, Eq.~\eqref{eq:bbjm}, and the two dots are coupled by the
  Maldacena--Qi bilinear, Eq.~\eqref{eq:H}. (b) Chord rules; solid and dashed
  chords are the chaotic ($n$) and integrable ($z$) species. In the disks phase no
  chord connects the two dots and the crossing weights are those of BBJM,
  Eq.~\eqref{eq:weights}. In the cylinder phase cross-boundary chords (gray) condense,
  and every chord---of either species, on either dot---carries the traveling penalty
  $e^{-\mu L}$ of Eq.~\eqref{eq:travel}, while the crossing weights are unchanged:
  $n$--$n$ and $n$--$z$ crossings weigh $q$, and $z$--$z$ crossings are trivial.
  (c) Why the penalty caps the order parameter: $|g_z|$, normalized to its plateau,
  against $\mu\tau$; the shaded strip is $\mu\tau\lesssim1$ and the dotted line is
  $\mu\tau=\mu\beta/2$. In the cylinder the source of Eq.~\eqref{eq:eom} is damped by
  $e^{-\mu\tau}$, so $g_z$ saturates within the strip at $2\kappa^2/\mu^2$; in the disk
  it keeps growing and reaches $\kappa^2\beta^2/4$ only at $\mu\beta/2$. Panel (c) is
  schematic and is drawn for $\mu\beta=55$.}
  \label{fig:model}
\end{figure}

For the coupled model we use the chord rules derived in
Ref.~\cite{Sontag:2026} for the MQ model and extended to the second species.
In the \emph{disks} phase ($\beta=\mathcal{O}(p^0)$) no chord connects the two
dots, and $Z=Z_{\rm BBJM}(\beta)^2$. In the \emph{cylinder} phase the
chords created by the bilinear condense, and two rules replace the free propagation:
\begin{enumerate}
  \item \textbf{Every} chord, of either species and on either dot, carries a
  traveling penalty
  \begin{equation}
    e^{-\ell(\tau_1,\tau_2)},\qquad
    \ell = \mu\min\!\big(|\tau_1-\tau_2|,\,\beta-|\tau_1-\tau_2|\big);
    \label{eq:travel}
  \end{equation}
  \item any two chords that \emph{must} cross carry the BBJM crossing weight
  for that pair of species, i.e.\ $e^{-\lambda}$ for $nn$ and $nz$; the $zz$
  crossing remains trivial, since commuting-SYK chords commute and the
  crossing is a property of the algebra rather than of the topology
  \cite{Sontag:2026}; we argue this at the chord-rule level in
  Appendix~\ref{app:penalty}.
\end{enumerate}
Rule 1 for the integrable species is not immediate, since the integrable chord
arises from a different contraction pattern; we verify it in
Appendix~\ref{app:penalty}. Rule 2 means that the BBJM crossing structure
\eqref{eq:weights} is unchanged; only the single-chord weight is modified.
Fig.~\ref{fig:model}(a,b) depicts the model and the two rules.

\section{The cylinder phase as a linear potential}
\label{sec:cylinder}

At the level of the chord rules and the semiclassical saddle, the BBJM
saddle is described by bilocal fields $g_a(\tau_1,\tau_2)$, with the
chord densities $n_a=-\frac12\partial_1\partial_2 g_a$ and the action of
Ref.~\cite{Berkooz:2024ofm},
\begin{equation}
\begin{split}
  S[n] = {}& \frac14\int_0^\beta\! d\tau_1\int_0^\beta\! d\tau_2
  \int_{\tau_1}^{\tau_2}\! d\tau_3\int_{\tau_2}^{\tau_1}\! d\tau_4\,
  \sum_{a,b=0}^{1}\M_{ab}\,n_a(\tau_1,\tau_2)\,n_b(\tau_3,\tau_4) \\
  &+ \frac12\int_0^\beta\! d\tau_1\int_0^\beta\! d\tau_2\,
  \sum_{a=0}^{1} n_a(\tau_1,\tau_2)
  \left[\ln\frac{n_a(\tau_1,\tau_2)}{\kappa_a^2J^2}-1\right] ,
\end{split}
  \label{eq:S}
\end{equation}
with $\kappa_0^2=\nu^2$ and $\kappa_1^2=\kappa^2$. Here $a=0$ is the chaotic
species and $a=1$ the integrable one, and we write $g_n\equiv g_0$,
$g_z\equiv g_1$ (and likewise $n_n\equiv n_0$, $n_z\equiv n_1$). For a single species
\eqref{eq:S} reproduces Eq.~(4.5) of Ref.~\cite{Berkooz:2024ofm} term by
term, including the ordering of the four nested integrations; we have not
altered the conventions of that reference. The traveling penalty
\eqref{eq:travel} weights each chord by $e^{-\ell}$, and since the density
$n$ counts chord \emph{endpoint pairs} (each chord contributing twice), it
adds
\begin{equation}
  S \;\longrightarrow\; S + \frac12\,\mu\int\!\!\int\sum_{a=0}^{1}
  n_a(\tau_1,\tau_2)\,\tau_{12},
  \label{eq:Spen}
\end{equation}
with $\tau_{12}=\min(|\tau_1-\tau_2|,\beta-|\tau_1-\tau_2|)$.

\emph{Which coupling?} A chord of length $L$ carries the $p$-th power of a
free propagator, and for $\mu\beta\gg1$ that propagator is
$\tfrac12 e^{-\eta L}$ with $\eta$ the coefficient of Eq.~\eqref{eq:H}. The
chord weight is therefore $(2\bar G)^p\sim e^{-p\eta L}\equiv e^{-\mu L}$:
\emph{the exponent of the traveling penalty is} $\mu=p\eta$, and $\eta$ itself
never multiplies a time difference. Every saddle in
Secs.~\ref{sec:cylinder}--\ref{sec:length} is a function of $\mu$, and $\mu$ is
the variable plotted in Figs.~\ref{fig:cap} and~\ref{fig:wl}. The topological
term and the phase criterion of Secs.~\ref{sec:phase}--\ref{sec:temperature}
carry $\eta$ instead, and it is there---and only there---that $\eta$ appears.

Varying \eqref{eq:S}+\eqref{eq:Spen} with respect to $n_a$, the
quartic term contributes $-\sum_b\M_{ab}g_b$ (the orientation of the
nested integrals in \eqref{eq:S} fixes its sign), the entropy term
contributes $\ln[n_a/(\kappa_a^2J^2)]$, and the penalty contributes
$+\mu\tau_{12}$. Stationarity gives
\begin{equation}
  n_a = \kappa_a^2J^2
  \exp\Big[\sum_b \M_{ab}g_b - \mu\tau_{12}\Big] .
  \label{eq:nsol}
\end{equation}
The penalty depends only on $\tau_{12}$, so translation invariance is intact
and the saddle still depends only on $\tau=\tau_1-\tau_2$. Using
$\partial_1\partial_2\to-\partial_\tau^2$ and the symmetry about $\beta/2$,
the equations reduce to a one-dimensional problem on $\tau\in[0,\beta/2]$,
where $\tau_{12}=\tau$:
\begin{equation}
  \boxed{\;
  g_a''(\tau) = 2J^2\kappa_a^2\,
  \exp\Big[\sum_{b=0}^{1}\M_{ab}\,g_b(\tau) - \mu\tau\Big]\;},
  \qquad g_a(0)=g_a'(\beta/2)=0 .
  \label{eq:eom}
\end{equation}
The boundary conditions follow from $g_a(\tau,\tau)=0$ at coincident
points and from the reflection symmetry $\tau\to\beta-\tau$, which makes
$\beta/2$ an extremum of $g_a$.

This is the central structural result of the paper: \textbf{the cylinder phase
of the coupled BBJM model is the BBJM model in a linear confining potential of
slope $\mu$.} The BBJM equations are unchanged; the only modification is the
factor $e^{-\mu\tau}$ multiplying the source. At $\mu=0$ Eq.~\eqref{eq:eom}
reduces to the disk equations, as it must.

The explicit $\mu\tau$ terms cancel between the entropy and the traveling
terms in \eqref{eq:S}+\eqref{eq:Spen}, so, after an integration by parts, the on-shell action
keeps its familiar form with the deformed solution,
\begin{equation}
  S_{\rm cyl} = \frac{\beta}{8}\int_0^\beta\! d\tau\,
  \sum_{a,b}\M_{ab}\,g_a'' g_b
  - \frac{\beta}{2}\int_0^\beta\! d\tau\sum_a \kappa_a^2J^2
  \exp\Big[\sum_b\M_{ab}g_b-\mu\tau\Big].
  \label{eq:onshell}
\end{equation}
We set $J=1$. We checked Eq.~\eqref{eq:onshell} against the known analytic
result of DSSYK, which is recovered at $\kappa=0$ \emph{and} $\mu=0$: for
$\beta J=110$ our solution gives $S=-215.43$ against
$-2\beta J+\pi^2/2=-215.07$, a $0.2\%$ difference attributable to the
$\mathcal{O}(\lambda)$ corrections that we drop.

\section{The cylinder caps long-chord order parameters}
\label{sec:cap}

The result of Sec.~\ref{sec:cylinder} has a consequence more general than the
integrable deformation that motivates it. In the cylinder phase the source of
Eq.~\eqref{eq:eom} is damped by $e^{-\mu\tau}$ on $\tau\in[0,\beta/2]$, so a
field cannot grow beyond a distance $\mu^{-1}$ from the boundary. For
$\mu\beta\gg1$ the source is confined to a boundary layer, and balancing
$g_a''\simeq2\kappa_a^2e^{-\mu\tau}$ with the boundary conditions
$g_a(0)=g_a'(\beta/2)=0$ gives
\begin{equation}
  \bigl\lvert g_a(\beta/2)\bigr\rvert \lesssim \frac{2\kappa_a^2}{\mu^2},
  \qquad \mu\beta\gg1 ,
  \label{eq:capgeneral}
\end{equation}
to be compared with the $\kappa_a^2\beta^2/4$ that a disk of the same
temperature allows for a field of the same coupling. Their ratio is
$(\mu\beta)^2/8$, which is large in the cylinder phase, and the cap is
correspondingly smaller by that factor (Fig.~\ref{fig:model}(c)). Operationally:
\emph{any saddle whose order parameter requires the field to grow like
$\beta^2$ over the whole circle is unavailable in the cylinder for
$\mu\beta\gg1$.} The remainder
of this section exhibits two such saddles---one for each of the two ways in
which the deformation can act---and verifies \eqref{eq:capgeneral} in each.

\subsection{Instance one: the quasi-integrable phase}
\label{sec:capinst1}

In the disk phase the broken saddle freezes the chaotic chords,
$g_n\simeq0$, and the active integrable field obeys $g_z''=2\kappa^2e^{g_n}
\simeq2\kappa^2$, so
\begin{equation}
  g_z(\tau) = \kappa^2\tau(\tau-\beta),\qquad
  g_z(\beta/2) = -\frac{\kappa^2\beta^2}{4} .
  \label{eq:gzdisk}
\end{equation}
In the cylinder phase the source is damped by $e^{-\mu\tau}$, so $g_z$ cannot
grow over the whole circle. For $\mu\beta\gg1$ the source is confined to a
boundary layer of width $\mu^{-1}$, and balancing $g_z''\simeq2\kappa^2
e^{-\mu\tau}$ with $g_z(0)=g_z'(\beta/2)=0$ gives
\begin{equation}
  g_z(\beta/2)\;\longrightarrow\;-\frac{2\kappa^2}{\mu^2},\qquad \mu\beta\gg1 .
  \label{eq:gzcap}
\end{equation}
The order parameter is therefore capped: the quasi-integrable phase needs
chords of length $\mathcal{O}(\beta)$, and the linear potential forbids
anything longer than $\mathcal{O}(\mu^{-1})$.

\begin{figure}[t]
  \centering
  \includegraphics[width=\textwidth]{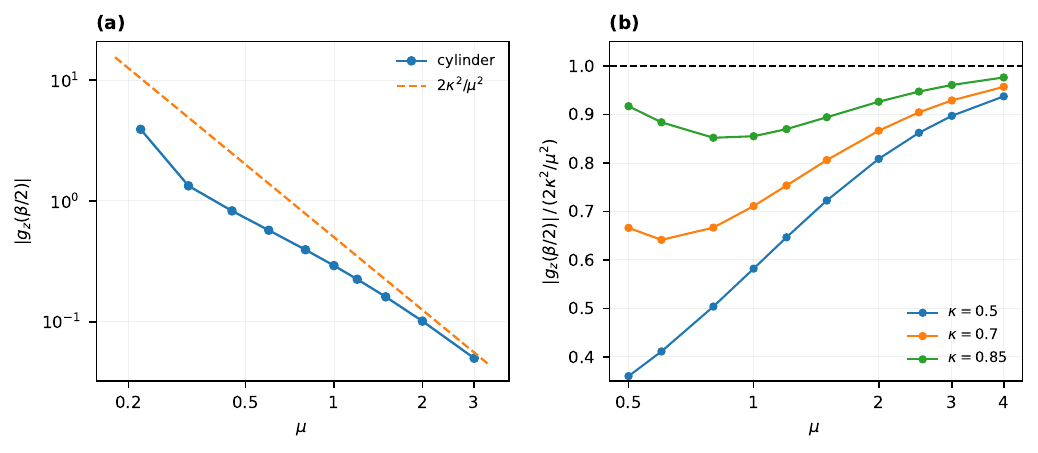}
  \caption{The quasi-integrable order parameter in the cylinder phase.
  (a) $|g_z(\beta/2)|$ against the exponent $\mu=p\eta$ of the traveling
  penalty, at $\beta J=110$ and $\kappa=0.5$ (points, from
  Eq.~\eqref{eq:eom}, obtained by continuation in $\mu$); the dashed line is
  the cap \eqref{eq:gzcap}, $2\kappa^2/\mu^2$, which the points approach from
  below as $\mu$ grows. In the disk phase the same quantity would be
  $\kappa^2\beta^2/4=756$, far off the top of the range shown. (b) The ratio
  $|g_z(\beta/2)|/(2\kappa^2/\mu^2)$ for $\kappa=0.5$, $0.7$ and $0.85$: all
  approach unity as $\mu$ grows, with the deviation ordered by $\nu^2$, as
  \eqref{eq:gzasy} requires.}
  \label{fig:cap}
\end{figure}

Fig.~\ref{fig:cap} shows the numerical solution of Eq.~\eqref{eq:eom} at
$\beta J=110$, $\kappa=0.5$, together with \eqref{eq:gzcap}; panel (b) shows
the same ratio for three couplings, which approaches unity as $\mu$ grows,
with the approach ordered by $\nu^2$. The crossover
between \eqref{eq:gzdisk} and \eqref{eq:gzcap} occurs at $\mu\beta\sim2\sqrt2$.
The bound \eqref{eq:capgeneral} is not tight everywhere---at $\mu=1.3$ its
right-hand side is $0.296$ while the saddle gives $0.20$---and it is worth
identifying the single reason why.

In the one-flavor case $M_{zz}=0$ closes the $z$-component of \eqref{eq:eom},
\begin{equation}
  g_z''=2\kappa^2J^2\,e^{\,g_n-\mu\tau}
  \quad\Longrightarrow\quad
  g_z(\beta/2)=-2\kappa^2J^2\!\int_0^{\beta/2}\!\!ds\;s\,e^{-\mu s}e^{g_n(s)} ,
  \label{eq:gzint}
\end{equation}
so $g_z$ never backreacts on itself, and the entire correction to
\eqref{eq:capgeneral} comes from the residual chaotic field $g_n$. That field
is itself capped, $\lvert g_n\rvert\le2\nu^2/\mu^2$ by the same argument, with
leading order $g_n^{(1)}(s)=-(2\nu^2/\mu^2)(1-e^{-\mu s})$. Inserting this in
\eqref{eq:gzint} and using
$\int_0^\infty s\,e^{-\mu s}ds=\mu^{-2}$ and
$\int_0^\infty s\,e^{-2\mu s}ds=(4\mu^2)^{-1}$ gives
\begin{equation}
  \bigl\lvert g_z(\beta/2)\bigr\rvert
  =\frac{2\kappa^2}{\mu^2}
  \Bigl[1-\frac32\,\frac{\nu^2}{\mu^2}
  +\mathcal{O}\bigl(\mu^{-4}\bigr)\Bigr] .
  \label{eq:gzasy}
\end{equation}
The cap is therefore an asymptotic \emph{equality}, and the parameter that
controls it is $\mu\gg\nu,\kappa$---which is \emph{not} the same condition as
$\mu\beta\gg1$. The boundary-layer picture of this section needs the latter;
the correction \eqref{eq:gzasy} needs the former. At $\mu=1.3$ one has
$\nu^2/\mu^2=0.44$, so the expansion does not apply there and the loose bound
is not a failure of \eqref{eq:gzasy}.

\begin{figure}[t]
  \centering
  \includegraphics[width=0.60\textwidth]{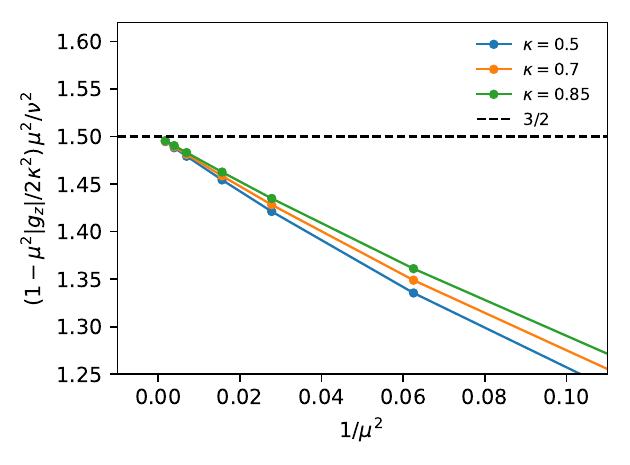}
  \caption{Test of the asymptotic form \eqref{eq:gzasy} of the cap. The
  quantity $\bigl(1-\mu^2\lvert g_z(\beta/2)\rvert/2\kappa^2\bigr)\mu^2/\nu^2$
  is plotted against $1/\mu^2$, for $\kappa=0.5$, $0.7$ and $0.85$ at
  $\beta J=110$. All three approach a common curve, which extrapolates to
  $3/2$ as $1/\mu^2\to0$, as \eqref{eq:gzasy} predicts: the leading
  correction to the cap depends on the couplings only through
  $\nu^2/\mu^2$, with a coefficient independent of $\kappa$.}
  \label{fig:capasy}
\end{figure}

Fig.~\ref{fig:capasy} verifies \eqref{eq:gzasy}. The combination
$\bigl(1-\mu^2\lvert g_z\rvert/2\kappa^2\bigr)\mu^2/\nu^2$ should tend to
$3/2$ as $1/\mu^2\to0$; it reaches $1.495$ at $\mu=24$, and the approach follows
a common curve for $\kappa=0.5$, $0.7$ and $0.85$, as \eqref{eq:gzasy} requires
since the leading correction depends on the couplings only through
$\nu^2/\mu^2$.

The free-energy consequence follows the same pattern. In the disk phase the
broken saddle gains $S_{\rm br}\simeq-\kappa^2\beta^2/2$ per dot over the
chaotic one (the saddle \eqref{eq:gzdisk} inserted in \eqref{eq:onshell}
at $\mu=0$), which at $\beta J=110$ and $\kappa=0.5$ is a difference of order
$10^3$. In the cylinder phase the corresponding gain is at most of order
$g=2\kappa^2/\mu^2$, and for most parameters it is zero because the broken
saddle has ceased to exist. At $\mu=0.1$, $\kappa=0.5$ we find a broken
cylinder saddle with $S_{\rm br}=-290.98$ against $-191.26$ for the chaotic
one, a gain of $100$ for $g=50$, i.e.\ twice $g$; at $\mu=0.5$, $\kappa=0.5$
seeds of every amplitude converge to the single unbroken solution with
$S=-123.30$, so there is no gain at all. The quasi-integrable phase therefore
barely competes in the cylinder, and over most of the parameter space it does
not compete at all. Note that $S_{\rm cyl}$ itself is \emph{not}
$-2\kappa^2/\mu^2$: the cylinder free energy is dominated by terms that are
present even at $\kappa=0$, and only the broken-minus-chaotic
\emph{difference} scales like $g$.

\subsection{Instance two: flavor symmetry breaking}
\begin{figure}[t]
  \centering
  \includegraphics[width=0.60\textwidth]{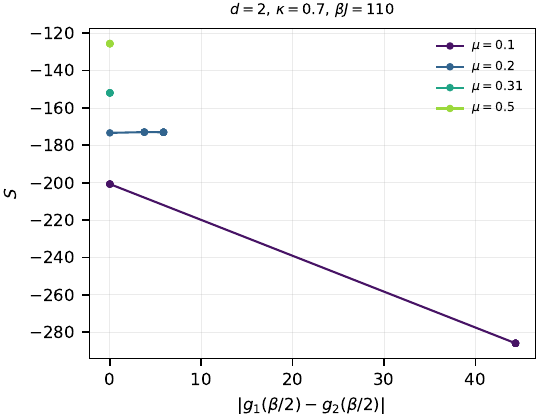}
  \caption{Free-energy profile for $d=2$ at $\kappa=0.70$ and $\beta J=110$:
  the action of every converged solution against its flavor asymmetry
  $\lvert g_1(\beta/2)-g_2(\beta/2)\rvert$, for twelve seed amplitudes
  $\in[0.01,2.0]$. At $\mu=0.10$ a strongly broken minimum exists and is
  lower by $85$; at $\mu=0.20$ it survives but has just been overtaken (its
  action is $0.33$ higher than the symmetric one); at $\mu=0.31$ and
  $0.50$ all seeds collapse to zero asymmetry, so no broken saddle exists.}
  \label{fig:flavprof}
\end{figure}

\label{sec:flavors}

The model \eqref{eq:H} carries a single integrable flavor per dot. With $d$
flavors, $\kappa_a^2=\kappa^2/d$ for $a\ge1$---the $d$-commuting construction of
Refs.~\cite{Gao:2024lve,Dai:2025multiflavor}---and the intersection matrix is
the one of Ref.~\cite{Dai:2025multiflavor}, namely $\M_{ab}=1$ for $a\ne b$
and $\M_{aa}=0$ for $a\ge1$ with $\M_{00}=1$; for $d=2$ this is
$\M=\begin{pmatrix}1&1&1\\1&0&1\\1&1&0\end{pmatrix}$ in the order
$(n,z_1,z_2)$. The distinction between $\M_{zz}=0$ for a single flavor and
$\M_{z_iz_j}=1$ for $i\ne j$ is the following. In the $d$-commuting
construction of Ref.~\cite{Dai:2025multiflavor} each $H_a$ is itself a
commuting-SYK Hamiltonian, but different copies do \emph{not} commute with
one another, because their fermion bilinears overlap. That is why $z_i$ and
$z_j$ chords cross with weight $q$, while the fermions of a single $z_i$
chord are locked into fixed bilinears and its crossing with itself is
trivial---the case treated in Appendix~\ref{app:penalty}. The two statements
are therefore consistent: $z_iz_i$ crossings are locked, $z_iz_j$ crossings
are not. The traveling penalty is unchanged, since it is
a property of the single-chord weight and not of the species
(Appendix~\ref{app:penalty}). The cylinder equations are therefore the same as
\eqref{eq:eom} with the multi-flavor $\M$.

We have solved these at $d=2$, $\beta J=110$, comparing the symmetric saddle
(all flavors equal) with the one-active-flavor saddle, from a wide range of
initial amplitudes. The flavor-broken saddle is much harder to reach in the
cylinder than in the disk. It exists only for weak $\mu$ and large $\kappa$,
and at $\mu=0.35$ and $\mu=0.50$ we find no broken saddle at all up to
$\kappa=0.85$; the symmetric solution is the only one. Where it does exist it
wins, with the lower action. The threshold grows quickly with $\mu$,
lying between $\kappa=0.40$ and $0.55$ at $\mu=0.10$ and between $0.70$ and
$0.85$ at $\mu=0.20$.

That the disappearance is a genuine loss of the broken saddle, rather than a
failure of the solver, is shown by the free-energy profile in
Fig.~\ref{fig:flavprof}. For each $\mu$ we scan twelve seed amplitudes
spanning two orders of magnitude and plot the action of every converged
solution against its flavor asymmetry. At $\mu=0.10$ the broken solution has
asymmetry $44.4$ and action $-285.97$ against $-200.74$ for the symmetric one;
at $\mu=0.20$ a broken solution still exists but its action is now $0.33$
\emph{higher} than the symmetric one, so the first-order flavor transition has
just taken place and the two saddles are nearly degenerate there; at $\mu=0.31$ and $0.50$ \emph{every} seed,
including the strongly asymmetric ones, converges to zero asymmetry. The
broken minimum has therefore ceased to exist, and it does so between
$\mu=0.20$ and $0.31$.

Bisecting in $\mu$ at fixed $d=2$ locates the disappearance of the broken
saddle. The value depends on the window of $\kappa$ that is probed: checking
$\kappa=0.60$, $0.70$ and $0.85$ at each step gives $\mu^{*}=0.31$, whereas
checking $\kappa=0.40$, $0.50$ and $0.60$ gives $\mu^{*}=0.14$. The ratio
$\mu^{*}/\langle\kappa\rangle^2$ is nearly the same in the two windows ($0.60$
and $0.56$), which suggests a criterion of the form $\kappa^2/\mu\gtrsim1.7$
rather than a fixed $\mu$; two windows are not enough to fix the exponent, and
we therefore quote
\begin{equation}
  \mu^{*}=0.31\quad(\kappa=0.60\text{--}0.85),
  \label{eq:mustar}
\end{equation}
i.e.\ the threshold for the window used throughout this paper, rather than a
universal constant. Above it the cylinder phase is flavor-symmetric at every
$\kappa$ we have probed.

Two remarks on the strength of that statement. First, the boundary is
located by \emph{continuing} the broken branch in $\mu$ rather than only by
scanning initial amplitudes: starting from $\mu=0.06$ and stepping in
$\Delta\mu=0.005$, one solution feeding the next, the asymmetry at
$\kappa=0.70$ decreases monotonically---$111.6$ at $\mu=0.06$, $44.4$ at
$0.10$, $17.9$ at $0.15$, $5.9$ at $0.20$---and the branch is lost at
$\mu=0.205$, with the residual jumping from $10^{-11}$ to $10^{-4}$. The
continuation therefore distinguishes `the branch ends' from `the solver
failed', which a scan over seeds alone cannot do. What it does not give is an
analytic bound.

Second, the broken saddle turns out to be sensitive to how
the chaotic field is seeded: it is reached from initial data in which $g_n$
starts at zero, but not from data in which $g_n$ starts at the DSSYK
profile. We therefore ran both classes of seed. At $\mu=0.10$ the broken
saddle appears in both---asymmetry $44.4$ and action $-285.97$ against
$-200.74$ for the symmetric one---while at $\mu\ge0.25$ both classes
converge to zero asymmetry, so the absence is not an artefact of the
seeding. At $\kappa=0.70$ the continuation puts it at $\mu^{*}=0.205$. The
loss of the branch is not a linear instability of the symmetric saddle: the
flavor-antisymmetric mode is stable at every $\mu$ we examined, so this is
not a pitchfork of that saddle, and the linear zero-mode condition cannot be
used to locate it.

The same asymmetries track the cap:
$2\kappa_a^2/\mu^2=136$, $49.0$, $21.8$ and $12.3$ at the same four values
of $\mu$, so the asymmetry stays within a factor of about two of the cap; the
branch is lost once the cap falls to the scale of the asymmetry the saddle
requires.

Two lessons follow. First, the wormhole suppresses not only the single-flavor
chaos--integrability transition but flavor symmetry breaking itself: the
wormhole phase is flavor-symmetric over most of its range. Second, the
$\sqrt d$ law of the disk phase does transfer to the cylinder. Here
$R(d,\kappa)\equiv(-\sum_{a=0}^{d}g_a)(\beta/2)\big|_{\kappa}
-(-\sum_{a=0}^{d}g_a)(\beta/2)\big|_{0}$, the total chord-number response
of Eq.~\eqref{eq:dldisk} for $d$ flavors, so the natural expectation is that
the broken cylinder saddle at flavor $d$ coincides with the $d=1$ saddle at
$\kappa_{\rm eff}=\kappa/\sqrt d$, i.e.\ $R(d,\kappa)=R(1,\kappa/\sqrt d)$. This is what we find. At $\mu=0.10$ the
ratio is $1.013$ and $0.982$ for $\kappa=0.45$ and $0.55$, and at
$\mu=0.20$ it is $1.011$, $1.023$ and $0.988$ for $\kappa=0.45$, $0.55$ and
$0.65$; the agreement is within $2.5\%$ wherever both saddles exist. We have
repeated the comparison at $d=3$, which is the next case where the broken
saddle has a different number of active fields: at $\mu=0.10$,
$R(3,\kappa)/R(1,\kappa/\sqrt3)=0.988$ and $0.998$ at $\kappa=0.65$ and
$0.75$, where the response is large ($R=14.8$ and $27.0$), so the law holds to
better than $1.2\%$. At $\mu=0.20$ the broken saddle is marginal and $R$ is
itself of order $10^{-2}$, so the ratio there is noise and we do not quote it.
The $\sqrt d$ law therefore does transfer to the cylinder, and it does so for
$d>2$. What does not
transfer is the \emph{survival} of the broken saddle: for fixed $\kappa$ it
disappears as $\mu$ grows, and that suppression is
the content of the cap, not a failure of the rescaling.

\section{The wormhole shields the geodesic length}
\label{sec:length}

The results of Secs.~\ref{sec:cylinder}--\ref{sec:cap} concern the free energy.
A more direct probe of the geometry is the bulk geodesic length. In DSSYK the
thermal correlator of an operator of dimension $\Delta$ is
\begin{equation}
  \langle O(\tau)O(0)\rangle = \big\langle e^{-\Delta\,\ell(\tau)}\big\rangle,
  \qquad \ell(\tau) = \lambda\,\mathcal{N}(\tau),
  \label{eq:dict}
\end{equation}
where $\mathcal{N}$ is the number of Hamiltonian chords crossed by the matter
chord \cite{Lin:2022rbf,Lin:2023sym,Sontag:2026}. Since the chord density is $n_a=-\frac12\partial_1\partial_2g_a$, the
chord number is $\mathcal{N}=-\frac12(g_n+g_z)$ evaluated at $\beta/2$; in
the one-dimensional reduction this is the first moment
$\mathcal{N}=\int_0^{\beta/2}\tau\,(n_n+n_z)\,d\tau$ of the density. The
collective field therefore gives \emph{twice} the chord number. In the disk phase, for one flavor, this is
$\mathcal{N}_{\rm disk}=\kappa^2\beta^2/8$. For $d$ flavors the answer depends on
the phase: in the flavor-symmetric phase all $d$ flavors are active with
$\kappa_a^2=\kappa^2/d$, so the total is again $\kappa^2\beta^2/8$, unchanged;
in the flavor-broken phase only one flavor is active and the chord number is
smaller by a factor $d$. The quantity plotted and tabulated
below is $\delta(-g)=2\,\delta\mathcal{N}$, so that $\delta\ell=\lambda\,\delta\mathcal{N}
=\frac{\lambda}{2}\,\delta(-g)$; we use the collective-field variable because
it is the one the saddle directly returns. Unlike the phase
boundary, which is set by a difference of on-shell actions, $\ell$ is a
chord-sector quantity and is $\mathcal{O}(1)$ in the double-scaling limit at
low temperature, where $\lambda\beta^2J^2=\mathcal{O}(1)$; near the
Hawking--Page point, where $\beta J=\mathcal{O}(1)$, it is small. It is
therefore the natural
observable in which to look for the effect of integrability.

In the disk phase $g_z$ is given by Eq.~\eqref{eq:gzdisk}, and since
$\ell=\lambda\mathcal{N}=-\frac{\lambda}{2}(g_n+g_z)$,
\begin{equation}
  R_{\rm disk}\equiv(-g)(\beta/2)\big|_\kappa-(-g)(\beta/2)\big|_0
  \simeq \frac{\kappa^2\beta^2}{4},
  \label{eq:dldisk}
\end{equation}
Here $\big|_0$ denotes the $\kappa\to0$ chaotic saddle of the \emph{same}
coupled model, not the broken branch continued to $\kappa=0$: $R$ is the
change in the \emph{total} number of chords, and it therefore contains the
change in $g_n$ as well as the growth of $g_z$. In the disk phase $g_n$ is
frozen to zero in the broken saddle while at $\kappa=0$ it carries the
chaotic profile, and the numerical value below shows that the $g_n$
contribution is small compared with the $g_z$ term.

The quantity is enormous: at $\beta J=110$ and $\kappa=0.5$ the analytic
value is $756$, to be compared with $-g_z(\beta/2)=754.74$ at the numerical
saddle; once the $\kappa=0$ baseline $-g(\beta/2)=7.15$ is subtracted the two
sides give $R=749.1$ analytically and $R=747.6$ numerically, a $0.2\%$
difference. In terms of the chord number this is
$\delta\mathcal{N}=R/2=374$, and since $\ell=\lambda\mathcal{N}$ the correlator
\eqref{eq:dict} is suppressed by $e^{-374\lambda\Delta}$; for any
$\lambda\Delta$ of order one the phase is effectively empty. Throughout we
quote $R=\delta(-g)$, which is the quantity the saddle returns directly, and
convert with $\delta\ell=\frac{\lambda}{2}R$ when a statement about $\ell$ is
intended. In the cylinder phase the capped order parameter
\eqref{eq:gzcap} suggests a much smaller response, but the actual value must be
computed, since Eq.~\eqref{eq:gzcap} is only asymptotic.

\begin{figure}[t]
  \centering
  \includegraphics[width=\textwidth]{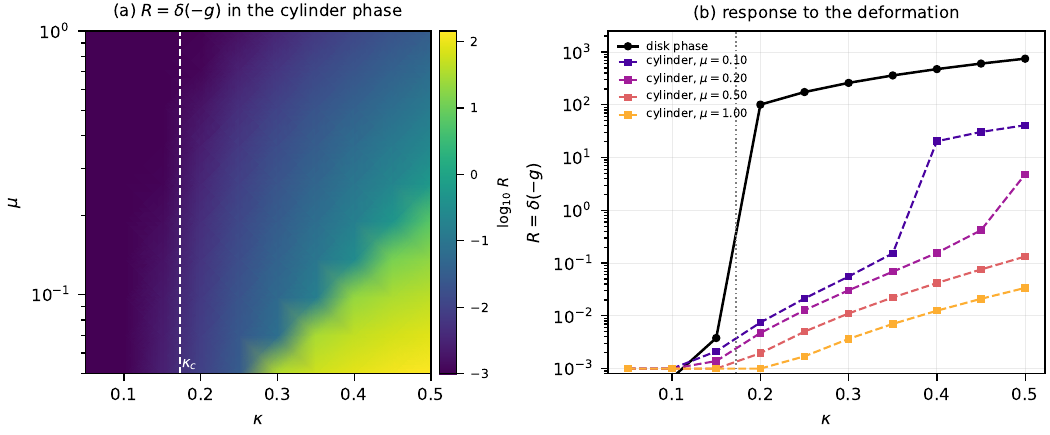}
  \caption{(a) The response of the geodesic length to the integrable
  deformation, $R=\delta(-g)$, in the cylinder phase, as a function of
  $\kappa$ and $\mu$ at $\beta J=110$ (color: $\log_{10}R$). The
  dashed line is $\kappa_c=0.1727$, the chaos--integrability threshold of
  Refs.~\cite{Berkooz:2024evs,Berkooz:2024ofm} expressed in the chord-model
  coupling (Appendix~\ref{app:penalty}) \emph{at $\beta J=110$}; the threshold
  moves with temperature (Sec.~\ref{sec:phase}). The dark region is where the wormhole is
  essentially transparent to the deformation. (b) Slices at fixed $\mu$
  (dashed, log scale) against the disk-phase response (solid). The disk phase responds at $\kappa_c$, by an
  amount $20$ to $2\times10^4$ times larger; the cylinder phase responds
  only at a much larger $\kappa$ and far more weakly. Below $\kappa_c$ the disk
  is chaotic and $R=0$, which appears here as the floor $10^{-3}$.}
  \label{fig:wl}
\end{figure}

Fig.~\ref{fig:wl} shows the result. In the cylinder phase $R$ is
smaller than in the disk phase, by a factor ranging from about $20$ to about
$2\times10^4$: at $\kappa=0.5$ and $\beta J=110$ it is $40.9$, $4.81$,
$0.133$ and $0.034$ for $\mu=0.10$, $0.20$, $0.50$ and $1.00$, against $748$ in
the disk phase. The correlator \eqref{eq:dict} is suppressed by
$e^{-(\lambda/2)R\Delta}$; for the disk value $R=748$ this is
$e^{-374\lambda\Delta}$, and for the cylinder values above it ranges from
$e^{-20\lambda\Delta}$ down to $e^{-0.017\lambda\Delta}$. Whether the
deformation destroys the phase therefore depends on $\lambda\Delta$, not on
$\Delta$ alone; the ratio of exponents between the two phases is what the
factors of $20$ to $2\times10^4$ above refer to.

The observable \eqref{eq:dict} is a same-boundary correlator. For the
wormhole a two-sided one is more natural, and it carries no further
information. In the cylinder phase the cross-boundary
geodesic length is related to the same-boundary one by a constant,
\begin{equation}
  e^{-\ell_{01}(\tau,\tau')}=e^{-\sigma}e^{-\ell_{00}(\tau,\tau')},
  \label{eq:ell01}
\end{equation}
with $\sigma$ the thickness of the tube \cite{Sontag:2026}. The two-sided
correlator is therefore proportional to \eqref{eq:dict},
$\langle O_LO_R\rangle\propto e^{-\Delta\sigma}
\langle O(\tau)O(0)\rangle$, with a proportionality constant that belongs to
the condensate and not to the integrable sector. The response
$R$ computed here is thus the response of the wormhole-crossing
length as well, up to that constant.

Two features of Fig.~\ref{fig:wl} deserve emphasis. First, the cylinder phase
has a \emph{transition of its own}: $R$ jumps sharply at a
$\kappa_w^{\rm cyl}(\mu)$ which is well above $\kappa_c$ and which grows as
$\mu$ decreases. This threshold is a property of the cylinder saddle itself and
is \emph{not} the thermodynamic boundary $\kappa_w$ of Sec.~\ref{sec:phase}.
Inside the cylinder the integrable sector remains inert up to that point. Second, even above it the response is bounded by
$O(100)$ rather than by $\kappa^2\beta^2/4$, so the wormhole never becomes
opaque to the deformation.

The physical reason is the one identified in Sec.~\ref{sec:cylinder}: the
linear potential starves the quasi-integrable saddle of the long chords it
needs. The wormhole does not so much \emph{resist} the integrable deformation
as leave it no room to develop, which is why both its thermodynamics
(Sec.~\ref{sec:phase}) and its geometry are almost blind to $\kappa$ until
$\kappa$ is large.

This connects to the real-time behavior of the same model without the
coupling. Reference~\cite{Aguilar-Gutierrez:2026jjv} solves the
Liouville equations of the BBJM model in real time and finds that the chord
number grows \emph{linearly} in the chaotic phase and \emph{quadratically} in
the quasi-integrable one, with a discontinuous change of growth law at the
transition; the Krylov complexity and the operator size both jump from exponential to
quadratic growth. Our $R$ is twice the same chord number evaluated at
imaginary time $\tau=\beta/2$, and the two descriptions agree: in the disk
phase $R=\kappa^2\beta^2/4$ is precisely the $\beta^2$ law that the
quasi-integrable saddle obeys, while in the cylinder the chemical potential of
Eq.~\eqref{eq:hcyl} cuts the growth off, so no $\beta^2$ law survives there.
The wormhole does not change which phase is which; it removes the quadratic
growth that a quasi-integrable saddle would otherwise produce.

\section{The phase diagram in the $(\kappa,\tilde\Delta)$ plane}
\label{sec:phase}

The results of Sec.~\ref{sec:length} concern the interior of the wormhole
phase. In this section we determine its boundary. The two phases compete
through their normalized free energies. In the
$\Tr\mathbf{1}=1$ normalization of Ref.~\cite{Berkooz:2024ofm},
\begin{equation}
  \ln Z_{\rm disks} = N\ln2-\frac{2S_{\rm BBJM}}{\lambda},\qquad
  \ln Z_{\rm cyl} = -\frac{S_{\rm cyl}}{\lambda}
  + N\ln\cosh\!\Big(\frac{\eta\beta}{2}\Big),
  \label{eq:lnZ}
\end{equation}
where the second term is the condensate contribution
$\ln Z_{\rm free}=N\ln2+N\ln\cosh(\eta\beta/2)$ minus the
$\ln\Tr\mathbf{1}=N\ln2$ of the normalization. The cylinder is favored iff
\begin{equation}
  D(\kappa,\mu)\;>\;\tilde\Delta,\qquad
  D\equiv 2S_{\rm BBJM}-S_{\rm cyl},\qquad
  \tilde\Delta\equiv -2p^2\Big[\ln\cosh\!\Big(\frac{\eta\beta}{2}\Big)-\ln2\Big],
  \label{eq:criterion}
\end{equation}
Both $D$ and $\tilde\Delta$ are pure numbers once $(\beta,\eta,\kappa)$ are
fixed, with $\mu=p\eta$ in $D$; at fixed $\eta$, $\tilde\Delta$ is a function of
temperature alone, so
it plays the role
of the Hawking--Page tuning parameter.

\begin{figure}[t]
  \centering
  \includegraphics[width=\textwidth]{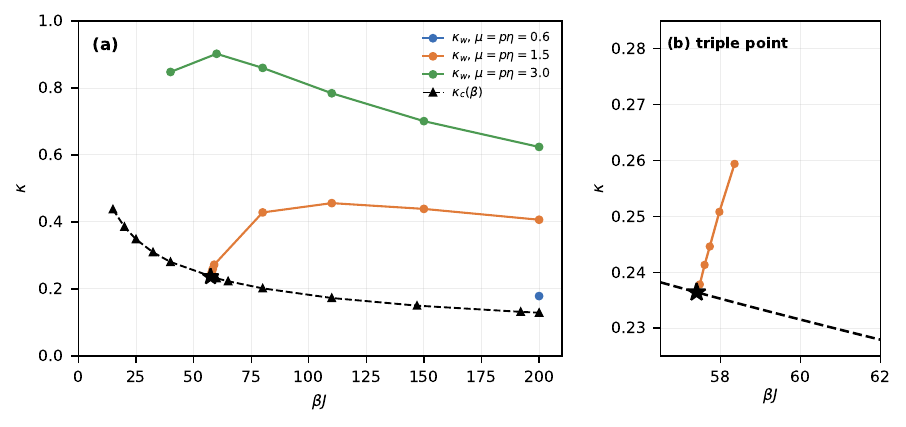}
  \caption{The wormhole boundary $\kappa_w$ (colored curves) and the
  chaos--integrability threshold $\kappa_c$ (black), both as functions of
  $\beta J$ so that the two can be compared at the same temperature. The
  thresholds are obtained from the crossing of the two disk saddles. Curves
  are drawn for the three values of $\eta$ at which $\kappa_w$ falls inside
  the scanned window $\kappa<0.95$ (labeled by $\mu=p\eta=0.6$, $1.5$ and
  $3.0$); for $\eta\ge0.2$ the boundary lies above
  $0.95$ at every $\beta$, so within the scanned window the wormhole survives
  at every $\kappa$. Everywhere both quantities exist, $\kappa_w\ge\kappa_c$;
  the boundary is not monotonic in $\beta$. (b) Zoom on the triple point of the
  $\eta=0.05$ curve: the boundary reaches the $\kappa_c$ line at
  $\beta J\simeq57.4$ and $\kappa\simeq0.236$, i.e.\ $\tilde\Delta\simeq-190$.}
  \label{fig:phase}
\end{figure}

One caution is needed about $\kappa_c$. The chaos--integrability threshold of
Refs.~\cite{Berkooz:2024evs,Berkooz:2024ofm} is a \emph{line} in the
$(\kappa,T)$ plane, not a number: solving for the crossing of the two disk
saddles gives $\kappa_c\simeq0.39$, $0.28$, $0.23$, $0.20$ and $0.17$ at
$\beta J=20$, $40$, $60$, $80$ and $110$. The value $\kappa_c=0.1727$ used
throughout is the one at $\beta J=110$, and statements such as
$\kappa_w\ge\kappa_c$ must be read at that temperature. In
Fig.~\ref{fig:phase} the two are therefore plotted against $\beta J$ and
compared there, rather than against a fixed threshold.

A remark on units is needed here. Because
$\tilde\Delta=-2p^2[\ln\cosh(\eta\beta/2)-\ln2]$, its absolute scale is set by
$2p^2$, and $p$ is not fixed in this paper. The entries
$\lvert\tilde\Delta\rvert=250$, $500$, $1000$, $2000$ of
Table~\ref{tab:D} are the dimensionless values of $\lvert\tilde\Delta\rvert$
in the normalization $2p^2=1800$ used there; they are \emph{not} to be
multiplied by $2p^2$ again. The choice $p\simeq30$ is a representative finite-$p$
normalization, not a convention of the literature: Refs.~\cite{Sontag:2026}
and~\cite{Berkooz:2024ofm} work directly in the double-scaling limit and do
not assign $p$ a value, and the only place where a number is needed is the
scaling of $\tilde\Delta$. With $1/p\simeq3\%$ this fixes the absolute scale
of the horizontal axis to that accuracy.

How much of this section survives a change of $p$? The criterion
$D(\kappa,\beta,\mu)>\tilde\Delta(\beta,\eta,p)$ contains $p$ twice: through
$\mu=p\eta$ in $D$, and through the $p^2$ in $\tilde\Delta$. At fixed
$(\eta,\lvert\tilde\Delta\rvert)$ both effects are present, and the boundary
moves a lot: at $(0.05,500)$ we find
$\kappa_w=0.291,0.351,0.380,0.403$ at $\beta J=79.7,65.0,59.7,55.9$ for
$p=20,30,40,60$, and at $(0.05,1000)$ the same sequence gives
$0.309,0.419,0.481,0.540$ at $\beta J=105.2,76.8,66.5,59.0$. Everything in
this section is a statement at $2p^2=1800$, and it should be read as such.

The $p\to\infty$ limit is not this table. There
$2p^2[\ln\cosh(\eta\beta/2)-\ln2]$ dominates $D$ unless
$\ln\cosh(\eta\beta/2)=\ln2$, so the boundary collapses onto the universal
Hawking--Page temperature of Sec.~\ref{sec:temperature} for every $\kappa$, and
the whole $\kappa$ dependence of Table~\ref{tab:D} is an
$\mathcal{O}(\lvert D\rvert/2p^2)$ effect---of order one at $p\simeq30$, which
is why the table is not already universal. Read in the other direction this is
a consistency check: raising $p$ at fixed temperature pushes $\kappa_w$ up, as
it must if the boundary is to reach $\eta\beta_{\rm HP}=2\,{\rm arccosh}\,2$.
Within Table~\ref{tab:D}, $\kappa_w$ decreases toward $\kappa_c$ as
$\lvert\tilde\Delta\rvert$ decreases and reaches it at the triple point
$\tilde\Delta=D_0(\eta)$; for $\tilde\Delta>D_0(\eta)$ the disks phase wins at
every $\kappa$ and there is no wormhole. The range of $\kappa$ over which the
wormhole survives is the ratio of $\lvert D\rvert$ to the scale of
$\tilde\Delta$, not a separate fine-tuning.

We evaluate $D$ as follows. For each pair $(\eta,\lvert\tilde\Delta\rvert)$ we
first solve $\tilde\Delta(\eta,\beta)=\tilde\Delta$ for $\beta$, then evaluate
$D(\kappa,\beta;\mu=p\eta)$ on a grid $\kappa\in[0.02,0.95]$ with spacing $0.01$ at
\emph{that} $\beta$, taking the lower of the chaotic and broken branches in
each phase, and finally solve $D=\tilde\Delta$ for $\kappa_w$.
Table~\ref{tab:D} collects the results.

\begin{table}[t]
  \centering
  \begin{tabular}{ccccc}
    \toprule
    $\eta$ & $\kappa_w$ at $\lvert\tilde\Delta\rvert=250$ & $500$ & $1000$ & $2000$ \\
    \midrule
    0.02 & --- & 0.146 & 0.175 & 0.188 \\
    0.03 & --- & 0.215 & 0.256 & 0.277 \\
    0.05 & 0.272 & 0.351 & 0.419 & 0.454 \\
    0.10 & 0.541 & 0.693 & 0.828 & 0.901 \\
    0.20 & --- & --- & --- & --- \\
    \bottomrule
  \end{tabular}
  \caption{The wormhole boundary $\kappa_w$, obtained self-consistently: for
  each $\lvert\tilde\Delta\rvert$ we first solve
  $\tilde\Delta(\eta,\beta)=\tilde\Delta$ for $\beta$ and then find the
  $\kappa$ at which $D(\kappa,\beta;\mu=p\eta)=\tilde\Delta$ at \emph{that} $\beta$.
  The computation uses $\lambda N=2p^2=1800$ ($p\simeq30$); the HP check of
  Sec.~\ref{sec:temperature} is performed both at this $p$ and at larger
  values, and the $1/p^2$ scaling of the correction is exhibited there. A dash has two meanings. For $\eta=0.02$ and $0.03$ at
  $\lvert\tilde\Delta\rvert=250$ the point lies above the triple-point value,
  $\tilde\Delta>D_0(\eta)$, so the disks phase wins at every $\kappa$ and no
  wormhole exists. For $\eta=0.20$ the wormhole does exist at every entry, but
  $\kappa_w$ lies above the scanned window $0.95$. The chaotic and broken
  branches are compared at each $\kappa$ and the lower action is used.}
  \label{tab:D}
\end{table}

The entries of Table~\ref{tab:D} depend on $\lambda$ only through the
combination $\lambda N=2p^2=1800$: any $\lambda\ll1$ with
$N=1800/\lambda\gg p^2$ gives the same table, and the saddle-point
approximation used throughout requires precisely that range. Their $p$
dependence, discussed above, is a leading-order effect; Table~\ref{tab:D}
and Fig.~\ref{fig:phase} are therefore statements about $p\simeq30$, while
the $p\to\infty$ limit is the universal Hawking--Page temperature.

Three features of Table~\ref{tab:D} drive the physics.

First, at fixed $(\mu,\beta)$ the function $D(\kappa)$ is monotonically
decreasing, and it has a knee at $\kappa_c$: for $\kappa<\kappa_c$ both phases
are chaotic and $D$ varies only weakly---through $\nu^2=1-\kappa^2$ in the
source---while for $\kappa>\kappa_c$ the disk phase becomes
broken and $2S_{\rm BBJM}\simeq-\kappa^2\beta^2$ drops steeply, dragging $D$
with it. For $\mu=0.50$ and $\beta J=110$ the values are $D=-310.50$,
$-310.52$, $-310.60$, $-413.71$, $-666.69$, $-989.23$, $-1825.60$, $-2908.12$
and $-4232.17$ at $\kappa=0.05$, $0.10$, $0.15$, $0.20$, $0.25$, $0.30$,
$0.40$, $0.50$ and $0.60$: strictly decreasing, and almost constant across the
chaotic window $\kappa<\kappa_c$, where it changes by $0.1$ out of $310$.

Second, because $D$ is monotone, for each $\tilde\Delta<D_0$ the boundary is a single number $\kappa_w$, and the wormhole occupies $\kappa<\kappa_w$. The
self-consistent values are collected in Table~\ref{tab:D}. The boundary
$\kappa_w$ \emph{increases} with $\eta$: a stronger Maldacena--Qi coupling
binds the cylinder more tightly, so it survives to larger $\kappa$. Similarly,
deeper into the wormhole phase (larger $\lvert\tilde\Delta\rvert$) the
boundary increases, because there the cylinder is favored over a wider range
of $\kappa$.

Third, $\kappa_w\ge\kappa_c$ at every coupling examined, with a comfortable
margin for $\eta\gtrsim0.03$, so that the three
phases---disk--chaotic, disk--quasi-integrable and cylinder---meet at a
\emph{triple point}
\begin{equation}
  \big(\kappa,\tilde\Delta\big) = \big(\kappa_c,\,D_0(\eta)\big).
  \label{eq:triple}
\end{equation}
Because $\kappa_c$ itself depends on temperature, the comparison must be made
at the $\beta$ that each $\lvert\tilde\Delta\rvert$ corresponds to. Doing so
we find $\kappa_w=0.351$ against $\kappa_c=0.223$ at $(\eta,\lvert\tilde\Delta\rvert)=(0.05,500)$,
$0.419$ against $0.206$ at $(0.05,1000)$, $0.175$ against $0.132$ at
$(0.02,1000)$ and $0.693$ against $0.31$ at $(0.10,500)$. The wormhole
therefore survives into the quasi-integrable phase at each of these points,
with a comfortable margin.

We emphasise that comparing against the fixed value
$\kappa_c=0.1727$ suggests a marginal violation at $\eta=0.02$. Most of that
is an artefact of using the $\beta J=110$ threshold at other temperatures, and
once $\kappa_c$ is evaluated at the temperature that
$\lvert\tilde\Delta\rvert$ implies the violation disappears: at
$\lvert\tilde\Delta\rvert=500$ the boundary is $0.146$ against
$\kappa_c=0.143$. The margin is only $0.003$, however, so at $\eta=0.02$ the
survival is marginal and it becomes firm only for $\eta\gtrsim0.03$. The
survival of the
wormhole into the quasi-integrable phase is therefore conditional on the
Maldacena--Qi coupling not being too small.

The physical reading is that integrability and the wormhole are not simply
competing. At the triple point the two transitions coincide: the wormhole dies
exactly when the dots become quasi-integrable. The pure Hawking--Page point
$\tilde\Delta=0$, however, lies in the disks phase; the deformation shifts the
onset of the wormhole to the lower temperature at which
$\tilde\Delta=D_0(\eta)$ (for $\eta=0.05$, $D_0=-187.5$ at $\beta J=57.4$), so
the wormhole is more robust than chaos only once it is inside its own phase.
There, the deformation is far less effective than in the disk phase, because
the confining potential of Sec.~\ref{sec:cylinder} prevents the
quasi-integrable order parameter from developing at all. Fig.~\ref{fig:phase} shows the two quantities that the comparison involves:
the boundary $\kappa_w(\beta)$ for several $\eta$, and the threshold
$\kappa_c(\beta)$, both as functions of temperature so that they can be read
off at the same point. Because $\kappa_w(\beta)$ is not monotonic---it peaks
near $\beta J\simeq100$---for a fixed $\kappa$ below its maximum the wormhole
occupies an interval of temperatures rather than everything below a
threshold. Fig.~\ref{fig:phase}(b) zooms on the triple point of the $\eta=0.05$
curve: it reaches the $\kappa_c$ line at $\beta J\simeq57.4$ and
$\kappa\simeq0.236$, i.e.\ $\tilde\Delta\simeq-190$.

\section{The Hawking--Page temperature}
\label{sec:temperature}

With $Z=\int\mathcal{D}n\,e^{-S/\lambda}$ one has $\ln Z=-S_{\rm on}/\lambda$.
The two phases differ in their topological term as well as in their chord
action: the disks phase carries the topological term $N\ln2$ ($\chi=2$) while
the cylinder phase carries $N\ln\cosh(\eta\beta/2)$ ($\chi=0$)
\cite{Sontag:2026}. The argument of that logarithm is not arbitrary: at $J=0$
the model is free and its partition function can be computed exactly,
$\ln Z=N[\ln2+\ln\cosh(\beta c/2)]$ for a bilinear of coefficient $c$
(Appendix~\ref{app:j0}). Hence
\begin{equation}
  \ln Z_{\rm disks}=N\ln2-\frac{2S_{\rm BBJM}}{\lambda},\qquad
  \ln Z_{\rm cyl}=N\ln\cosh\!\Big(\frac{\eta\beta}{2}\Big)
  -\frac{S_{\rm cyl}}{\lambda},
\end{equation}
and the cylinder is favored iff
\begin{equation}
  \ln\cosh\!\Big(\frac{\eta\beta}{2}\Big)=\ln2-\frac{D}{2p^2},\qquad D<0,
  \qquad D\equiv 2S_{\rm BBJM}-S_{\rm cyl},
  \label{eq:hpexact}
\end{equation}
where $\lambda N=2p^2$.

In the double-scaling limit the second term on the right is
$\mathcal{O}(1/p^2)$ and the condition reduces to
$\ln\cosh(\eta\beta/2)=\ln2$, i.e.
\begin{equation}
  \eta\beta_{\rm HP}=2\mathrm{arccosh}(2)=2.6339 .
  \label{eq:hpvalue}
\end{equation}
This is the exact leading-order value, and it does not involve $\kappa$: the
Hawking--Page temperature is \emph{universal} at leading order, and the
deformation enters only through the $\mathcal{O}(\lvert D\rvert/p^2)$
correction in \eqref{eq:hpexact}. The $\lvert D\rvert$ there is evaluated
\emph{at} the Hawking--Page point, where $\eta\beta\simeq2.6$ and hence
$\lvert D\rvert$ is itself small---for $(\kappa,\eta)=(0.50,0.50)$ we find
$\beta J=5.27$ and $\lvert D\rvert=14.0$ (evaluated at $\mu=p\eta=15$), so
$\lvert D\rvert/(2p^2)=0.008$ at $p=30$. It is not the large
$\lvert D\rvert$ of the deep cylinder phase, where instead $\lvert D\rvert/(2p^2)$ is not small
compared with $\ln2$.

A different number circulates in the literature, and the reason is instructive. Expanding
$\ln\cosh x\simeq x-\ln2$ and equating the cylinder value $N\eta\beta/2$ with
$N\ln2$ gives $\eta\beta_{\rm HP}=2\ln2=1.3863$, smaller by a factor $1.9$.
The expansion is not applicable at its own solution: there
$x=\ln2=0.693$, where $\ln\cosh x=0.223$ against $x-\ln2=0$. Applied
\emph{consistently} to $\ln\cosh x=\ln2$ it would instead give
$4\ln2=2.773$, within $5\%$ of \eqref{eq:hpvalue}; the factor of $1.9$ comes
from using the approximation in one place and not the other.

\begin{figure}[t]
  \centering
  \includegraphics[width=0.62\textwidth]{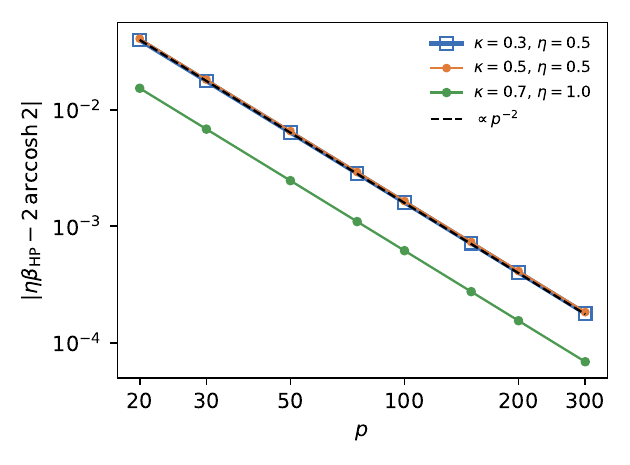}
  \caption{Finite-$p$ correction to the Hawking--Page temperature. The
  deviation $|\eta\beta_{\rm HP}-2\,{\rm arccosh}\,2|$ from the exact condition
  \eqref{eq:hpexact} is shown against $p$ for three $(\kappa,\eta)$ pairs; the
  dashed line is $\propto p^{-2}$.}
  \label{fig:hp}
\end{figure}

We have checked \eqref{eq:hpexact} numerically, evaluating $D$ at
$\mu=p\eta$. Solving it for $\beta$ at $(\kappa,\eta)=(0.30,0.50)$,
$(0.50,0.50)$ and $(0.70,1.00)$ gives $\eta\beta_{\rm HP}=2.6355$, $2.6356$
and $2.6345$ at $p=100$, against $2\mathrm{arccosh}(2)=2.6339$. The residual
is the $\mathcal{O}(1/p^2)$ correction, and it scales as predicted: for
$(\kappa,\eta)=(0.30,0.50)$ the deviation from $2\mathrm{arccosh}(2)$ falls
from $+0.0177$ at $p=30$ to $+0.0002$ at $p=300$, a factor of about $100$ as
$p^2$ requires. Fig.~\ref{fig:hp} shows the same deviation for three $(\kappa,\eta)$ pairs
over $p=20$--$300$: the points fall as $p^{-2}$ over more than a decade, the
three sets are parallel, and the two with $\eta=0.5$ nearly coincide. The leading-order value
is therefore accurate to $0.7\%$ already at
$p=30$, the value used for the phase diagram of Sec.~\ref{sec:phase}.

One consistency check tests the identification $\mu=p\eta$ of
Sec.~\ref{sec:cylinder}. The saddle description
requires $\mu\beta\gg1$, and $\mu=p\eta$, so at the Hawking--Page point
$\mu\beta_{\rm HP}=p\cdot2.6339\simeq79$ for $p\simeq30$: the transition sits
comfortably inside the boundary-layer regime in which Secs.~\ref{sec:cylinder}
--\ref{sec:cap} are derived. Reading $\eta$ instead as the exponent of the
traveling penalty would place the transition at $\eta\beta_{\rm HP}=2.63$,
i.e.\ at the edge of its own validity, and this is one of the reasons we keep
the two couplings apart throughout.

\section{Conclusions}
\label{sec:conclusions}

We have studied the Maldacena--Qi coupled double-scaled model in which each
dot carries a BBJM integrable deformation. The organizing result is a
statement about the phase rather than about the deformation: the cylinder
phase imposes a linear confining potential, and consequently caps the order parameter at $2\kappa_a^2/\mu^2$ in place of the disk value $\kappa_a^2\beta^2/4$,
so that any saddle requiring the field to grow like $\beta^2$ over the whole
circle is unavailable there for $\mu\beta\gg1$ (Sec.~\ref{sec:cap}).
Integrability is one such
saddle and flavor symmetry breaking is another. Our results are:

\begin{enumerate}
  \item The cylinder phase is, at the level of the chord rules and the
  semiclassical saddle, a BBJM chord model in a linear confining
  potential of slope $\mu=p\eta$. The traveling penalty of Ref.~\cite{Sontag:2026}
  depends only on the time separation, so the bilocal saddle collapses to one
  dimension and the BBJM equations survive with a damped source,
  Eq.~\eqref{eq:eom}.
  \item More generally the potential caps the order parameter, $|g_a(\beta/2)|\lesssim2\kappa_a^2/\mu^2$ against
  $\kappa_a^2\beta^2/4$ in the disk, so \emph{any} saddle whose order parameter
  requires the field to grow like $\beta^2$ over the whole circle is
  unavailable in the cylinder for $\mu\beta\gg1$ (Sec.~\ref{sec:cap}). For
  the quasi-integrable instance the bound is in fact the leading term of an
  asymptotic expansion, Eq.~\eqref{eq:gzasy}, whose first correction is fixed
  by the residual chaotic field and verified numerically in
  Fig.~\ref{fig:capasy}. We
  exhibit two instances:
  the quasi-integrable phase (Fig.~\ref{fig:cap}), which the wormhole geometry
  therefore cannot host, and flavor symmetry breaking
  (Sec.~\ref{sec:flavors}).
  \item Nevertheless $\kappa_w\ge\kappa_c$ at every coupling examined, with a
  comfortable margin for $\eta\gtrsim0.03$: the wormhole
  survives into the quasi-integrable phase and the three phases meet at a
  triple point (Fig.~\ref{fig:phase}). At the weakest coupling, $\eta=0.02$, the two are comparable: at $\lvert\tilde\Delta\rvert=1000$ the
  boundary is $\kappa_w=0.175$ against $\kappa_c=0.132$, while at
  $\lvert\tilde\Delta\rvert=500$ it is $0.146$ against $0.143$, a margin of
  only $0.003$ rather than a violation. The survival at $\eta=0.02$ is
  therefore marginal, and firm only for $\eta\gtrsim0.03$.
  \item The Hawking--Page temperature is fixed by the exact condition
  \eqref{eq:hpexact}, whose leading-order solution is the universal
  \eqref{eq:hpvalue}. At finite $p$ the condition is a self-consistency
  problem---$D$ itself depends on $\beta$---which we solve numerically, so the
  temperature tracks $\lvert D\rvert$ and hence the deformation. What is robust
  instead is the phase diagram in the $(\kappa,\tilde\Delta)$ plane, which is
  fixed by comparing $D$ with $\tilde\Delta$ (Sec.~\ref{sec:phase}).
\end{enumerate}

Several caveats should be stated. The most important concerns what the
word ``integrable'' means here. The deformation we use is the commuting-SYK
Hamiltonian, a specific integrable point whose spectrum is essentially
Gaussian; it is pseudo-holographic rather than a generic integrable system.
The principle of Sec.~\ref{sec:cap} is stated for long-chord order parameters
in general, but every quantitative statement in this paper is for this one
deformation, and a generic integrable deformation with a different chord
content could in principle behave differently.

The multi-flavor case is treated at $d=2$, and the
$\sqrt d$ law is checked up to $d=3$ (Sec.~\ref{sec:flavors}); a systematic
scan in $d$ is left for future work, as is a precise determination of the critical $\mu$ beyond which the
cylinder carries no flavor-broken saddle. We take the
lower of the chaotic and broken branches in each phase and do not perform a
full thimble decomposition. Where the analytic estimate
$\lvert D\rvert\simeq\kappa^2\beta^2$ is used it serves only to exhibit the
scaling of the boundary; every number quoted above is obtained from the
numerical $D$. Finally,
we use the coarse-grained chord action of Ref.~\cite{Berkooz:2024ofm} and
the cylinder rules of Ref.~\cite{Sontag:2026} rather than re-deriving them from
the microscopic $G,\Sigma$ action, and assume the crossing structure carries
over. The one ingredient the cylinder
description actually depends on---that integrable chords carry the same
traveling penalty as chaotic ones---we do derive, in
Appendix~\ref{app:penalty}. We also note that the pure MQ model has a third
saddle, a cylinder with a baby universe \cite{Sontag:2026}; Sec.~\ref{sec:phase}
compares only the disks and cylinder actions, and how the deformation
affects that saddle we leave open.

\section*{Acknowledgments}
We thank Shanghai University for support.

\appendix

\section{The integrable chord carries the traveling penalty}
\label{app:penalty}

The cylinder description of Sec.~\ref{sec:cylinder} requires that integrable
($z$) chords carry the same traveling penalty \eqref{eq:travel} as the
chaotic ones. We give a single argument covering both species. The point is
that the traveling penalty belongs to the \emph{free propagator} of the
coupled theory, and that propagator is common to all terms of the
Hamiltonian, because every term is built from the same fermions.

\subsection{Preliminaries: the chord action}
\label{app:coarse}

The chord action \eqref{eq:S} is the coarse-grained chord path integral of
Refs.~\cite{Berkooz:2024evs,Berkooz:2024ofm}; see also
Ref.~\cite{Dai:2026bulk} for a recent construction of bulk contact
interactions from chord diagrams built on Parisi's hypercube
\cite{Berkooz:2023parisi}. We sketch the chain below; the algebra is
in Ref.~\cite{Berkooz:2024ofm}. Divide the thermal circle into $s$ segments and let
$n_{ij}$ be the number of chords stretching from segment $i$ to segment $j$.
For fixed $\{n_{ij}\}$ the weight of a chord diagram is a product of four
factors: the amplitude $\langle n_i|e^{-\beta_iT}|0\rangle$ to generate $n_i$
chords from a segment, the $q$-multinomial that splits them into groups
$n_{ij}$, the $q$-factorials that reorder each group, and the crossing weight
$q^{\sum_{i<k<j<\ell}n_{ij}n_{k\ell}}$. Summing over $\{n_{ij}\}$ gives the
exact partition function
\begin{equation}
Z=\sum_{\{n_{ij}\}}q^{\sum_{i<k<j<\ell}n_{ij}n_{k\ell}}
\prod_i\frac{[n_i]_q!}{\sqrt{\prod_{j\neq i}[n_{ij}]_q!}}
\langle n_i|e^{-\beta_iT}|0\rangle ,
\label{eq:coarseexact}
\end{equation}
with $q=e^{-\lambda}$ and $[n]_q!$ the $q$-factorial; the square root reflects
that each pair of segments is counted twice. In the double-scaled
limit $\lambda\to0$ with $\tilde n_{ij}=\lambda n_{ij}$ held fixed, the
$q$-factorials exponentiate; the continuum limit $s\to\infty$ turns the sums
into integrals over $n(\tau_1,\tau_2)$ and gives \eqref{eq:S}. The extension
to two species is the replacement
$\sum_{i<k<j<\ell}n_{ij}n_{k\ell}\to\sum_{a,b}\M_{ab}n^a_{ij}n^b_{k\ell}$,
which is how $\M_{ab}$ and $\kappa_a^2J^2$ enter \eqref{eq:S}.

The MQ coupling adds one rule: every chord, of either species and on either
dot, carries the traveling weight $e^{-\mu L}$ of Eq.~\eqref{eq:travel}. In
\eqref{eq:coarseexact} this is a per-chord factor; in the continuum it adds
the symmetric term \eqref{eq:Spen}. Because the factor is blind to the species
and to the dot, the saddle equation is \eqref{eq:eom}, with the same
$e^{-\mu\tau}$ in the source of every species. This is the sense in which the
penalty is species-blind; the free-propagator argument and the explicit checks
are given below.

\subsection{The free propagator is a property of the coupling, not of the
species}

In \eqref{eq:H} only the Maldacena--Qi term is quadratic in the fermions; the
two Hamiltonian terms are $p$-local. Treating the latter as perturbations, the
Gaussian part of the path integral is fixed by the $\eta$ coupling alone, and
its propagator is the $2\times2$ matrix $G^{(0)}_{ab}(\tau,\tau')$ obtained by
inverting the quadratic form of Ref.~\cite{Sontag:2026}. This propagator
knows about $\eta$, $\beta$ and the two dots; it knows nothing about
$H_{\rm DSSYK}$ or $H_{\rm int}$. In the cylinder phase \cite{Sontag:2026}
\begin{equation}
  \big(2\bar G^{(0)}_{aa}(\tau,\tau')\big)^p
  = e^{-\mu\min(|\tau-\tau'|,\beta-|\tau-\tau'|)},
  \qquad a=L,R .
  \label{eq:gmup}
\end{equation}

\subsection{The fermionic content of a chord is common}

Expanding $e^{-\beta H}$ and averaging over the disorder, the couplings
$J_{i_1\dots i_p}$ of the DSSYK terms and $B_{i_1\dots i_{p/2}}$ of the
integrable terms are independent Gaussians, and the average factorizes,
$\langle\cdot\rangle=\langle\cdot\rangle_J\langle\cdot\rangle_B$. A chord is a
contraction of two couplings, so its weight contains $\langle J^2\rangle$ or
$\langle B^2\rangle$; since $\langle JB\rangle=0$ by independence, \emph{no
chord connects a DSSYK insertion to an integrable one}. A chord is therefore
either purely chaotic or purely integrable. This is what makes the
two-species description consistent, and it is a consequence of the disorder
structure rather than an assumption.

Let a chord join two insertions that both act on the same dot, so that every
fermion line it contains is a propagator $G^{(0)}_{LL}$ of that dot. Both
insertions are $p$-local---this is stated explicitly in
Ref.~\cite{Berkooz:2024ofm}, and for the commuting-SYK case the integrable
Hamiltonian is
\begin{equation}
  H_{\rm I\text{-}SYK}=i^{p/2}\!\!\sum_{1\le i_1<\dots<i_{p/2}}\!\!
  B_{i_1\dots i_{p/2}}
  \big(\psi_{2i_1-1}\psi_{2i_1}\big)\cdots
  \big(\psi_{2i_{p/2}-1}\psi_{2i_{p/2}}\big),
  \label{eq:isyk}
\end{equation}
a product of $p/2$ bilinears, i.e.\ $p$ fermions. The coupling contraction
therefore leaves $p$ fermions to be paired, and each of the $p$ lines is a
propagator $G^{(0)}_{LL}(\tau,\tau')$. How the $p$ fermions are \emph{grouped}
--- as $p$ separate fields, as in the DSSYK term, or as $p/2$ bilinears, as in
\eqref{eq:isyk}---changes only the combinatorial prefactor, not the propagator
content: the chord brings $[G^{(0)}_{LL}(\tau,\tau')]^p$ times a constant.

For the chaotic chord this is immediate: the $p$ lines pair freely, giving
$\big[2G^{(0)}_{LL}\big]^p$, which is the factor appearing in the MQ expansion
\cite{Sontag:2026}. For the integrable chord the coupling contraction forces
the two insertions to carry the same set $\{i_k\}$, so the pairing is
constrained to be bilinear by bilinear,
\begin{equation}
  \big\langle b_k(\tau)b_k(\tau')\big\rangle
  =-\big\langle\psi_A(\tau)\psi_A(\tau')\big\rangle
   \big\langle\psi_B(\tau)\psi_B(\tau')\big\rangle
  =-G^{(0)}_{LL}(\tau,\tau')^2 ,
  \label{eq:bilin}
\end{equation}
and
\begin{equation}
  \Big[-G^{(0)}_{LL}{}^2\Big]^{p/2}
  =(-1)^{p/2}\big[G^{(0)}_{LL}\big]^{p}
  =(-1)^{p/2}2^{-p}\big[2G^{(0)}_{LL}\big]^{p} .
  \label{eq:intpow}
\end{equation}
The sign and the factor $2^{-p}$ are independent of $\tau$ and are absorbed
into the normalization of $\kappa$. In both cases the $\tau$ dependence is the
same $p$-th power of the same propagator. An explicit check for $p=2,4,6$ with
the exact free MQ propagator gives the weights $(2G)^p$ and $(-G^2)^{p/2}$,
hence a ratio $(-1)^{p/2}2^{-p}$ for every $\tau$ and the same decay
$e^{-p\eta L}$. Conversely, solving the saddle with the penalty applied only
to the chaotic species does not produce a capped solution: the $z$-field is
undamped and runs away. The cap is therefore a consequence of the penalty
being species-blind, not of the cylinder geometry alone.

The absorption is a definition, so we state exactly what is defined. In Eq.~\eqref{eq:H} $\kappa$ is the coupling of the \emph{chord
model}, i.e.\ the coefficient that appears in the collective action
\eqref{eq:S} and in Eq.~\eqref{eq:eom}; it is not the variance of the
microscopic couplings $B_{i_1\dots i_{p/2}}$. At equal microscopic variance
the integrable chord weight is $2^{-p}$ times the chaotic one, so the ratio
of the chord-model couplings is
\begin{equation}
  \frac{\kappa^2}{\nu^2}
  =\frac{B^2}{J^2}\,2^{-p}
  \label{eq:kappaB}
\end{equation}
up to the sign $(-1)^{p/2}$; equivalently, a chord-model coupling of order
one corresponds to $B^2/J^2=\mathcal{O}(2^p)$. Because $p$ is sent to infinity this factor
cannot be absorbed into a $p$-independent number, and it is therefore part of
the definition of the chord-model coupling rather than a numerical
coincidence. Two consequences follow. First, the threshold $\kappa_c$ (linear
in $\kappa$) and the cap $2\kappa^2/\mu^2$ (quadratic) are statements about
the chord-model coupling; expressed in terms of the microscopic variance they
rescale by $2^{-p/2}$ and $2^{-p}$ respectively, which changes no physical
statement. Second, any comparison with a microscopic (finite-$N$)
determination of $\kappa$ must first translate through this factor.

\subsection{Why $q_{zz}=1$ survives in the cylinder}

Rule 2 of Sec.~\ref{sec:model} replaces free propagation by a topological
arrangement in which certain chords are forced to cross. It is important that
this does not change the \emph{weight} of a crossing, only its multiplicity.

The weight $q=e^{-\lambda}$ per crossing is a ratio of structure constants of
the chord algebra: each crossing contributes one factor of $q$ because the
Hamiltonian chords generate $\mathrm{U}_q(\mathfrak{sl}_2)$
\cite{Berkooz:2018jqr,Lin:2022rbf}. Two integrable chords do not generate such
a factor. The value $M_{zz}=0$ is derived from the microscopic contraction in
the appendix of Ref.~\cite{Berkooz:2024ofm}: there the overlap of two
integrable index sets is counted, and because the corresponding operators
commute, the $zz$ overlap contributes a factor of one, whereas the $nn$ and
$nz$ overlaps are Poisson distributed and give $q$ and
$q_{nz}=e^{-2p\,p_{\rm Integ}/N}$ respectively, with $p_{\rm Integ}$ the body
size of the integrable term (here $p_{\rm Integ}=p$, so $q_{nz}=e^{-\lambda}=q$). Equivalently, the fermions of a $z$-chord are locked
into $p/2$ bilinears on fixed pairs and the crossing cannot permute them. The
algebra generated by $H_{\rm int}$ is commutative, so the corresponding ratio
of structure constants is one. This is a property of the species, not of how the chords are
arranged on the thermal circle. A forced crossing of two $z$-chords in the
cylinder is therefore still weighted by one, while a forced $nn$ or $nz$
crossing still carries $e^{-\lambda}$. Hence the intersection matrix (3) is
unchanged, which is what Eq.~\eqref{eq:eom} requires.

The decisive point is that commutativity is a property of the \emph{microscopic}
Hamiltonian and not of the phase. For any fixed realization of the couplings
$B_{i_1\dots i_{p/2}}$, the terms of $H_{\rm int}$ commute with one another
identically, because each is a product of bilinears $\psi_{2i-1}\psi_{2i}$ on
disjoint pairs and the bilinears of a given pair commute with everything in
$H_{\rm int}$. This holds before any disorder average and before any
semiclassical limit (we have also verified
$[H_{\rm int}(B),H_{\rm int}(B')]=0$ numerically for random couplings), so the
chord algebra generated by $H_{\rm int}$ is commutative for every configuration---hence $q_{zz}=1$ cannot depend on which
saddle of the coupled model one expands about. Rewriting $q_{zz}=e^{-\lambda}$
in the cylinder would amount to a $q$-deformation of an algebra that is exactly
commutative microscopically, which is not possible.

The extension to the cylinder can be phrased in the method of
Ref.~\cite{Sontag:2026}, who derive their rules by solving the free $G,\Sigma$
theory exactly and then expanding the partition function in powers of the
coupling. In the present model the free theory is still the quadratic $\eta$
term alone, and its propagator $G^{(0)}_{ab}$ is built from the same fermions
for every species---it carries the geometry (the $\min$ structure of the
cylinder) but no species label. The chord weight $(2G^{(0)}_{ab})^p$ is
therefore common to $n$- and $z$-chords, which is the content of
Appendix~\ref{app:penalty}. The crossing weights, by contrast, come from the
combinatorics of the fermionic contractions and depend on the species---on
whether the operators commute---but not on the propagator. Geometry enters
only through $G^{(0)}$ and species only through the contractions, so the two
cannot mix: a forced crossing in the cylinder changes how many crossings
there are, not what each one weighs. This is why $M_{zz}=0$ survives, and it
is a consequence of the factorization rather than an extra assumption.

What the cylinder does change is which crossings are \emph{forced}: chords
wrapping the tube must cross chords on the other side. That is a statement
about the arrangement, and it multiplies the number of crossings, not their
weights. This is why the intersection matrix (3) is unchanged and
Eq.~\eqref{eq:eom} keeps $M_{zz}=0$.

Combining \eqref{eq:intpow} with \eqref{eq:gmup}, the integrable chord is
weighted by $e^{-\mu\min(|\tau-\tau'|,\beta-|\tau-\tau'|)}$ up to a constant,
exactly as the chaotic one. Equivalently, in the collective field both species
feel a common additive term $-\mu\tau$ in the exponent, independent of the
species index $a$, which is the structure of Eq.~\eqref{eq:eom}.

\subsection{The cylinder as a chemical potential for chord number}

There is a second and simpler way to state the same result, following the
chord-Hamiltonian formulation of Ref.~\cite{Sontag:2026}. Slicing a cylinder
diagram at $\tau=0$ and $\tau=\beta/2$ gives a chord Hilbert space whose
Hamiltonian carries, in addition to the transfer matrix of the chord algebra,
a term proportional to the total chord number,
\begin{equation}
  H^{\rm cyl}_A = -\frac{J}{\sqrt\lambda}\,T^{\rm cyl}_A
  + \mu\,N_A ,
  \label{eq:hcyl}
\end{equation}
with $N_A$ the number of chords and $\mu$ the coefficient of the traveling
penalty. In this language the penalty is not a modification of the propagation rule
but a term linear in the chord number; for a chord of span $L$ the weight
$e^{-\mu L}$ is the Boltzmann factor of a chemical potential conjugate to
the chord's length, and the operator $N_A$ counts the chords it acts on.

For two species \eqref{eq:hcyl} reads
\begin{equation}
  H^{\rm cyl} = -\frac{\nu}{\sqrt\lambda}T_n
  -\frac{\kappa}{\sqrt\lambda}T_z + \mu\,(N_n+N_z),
\end{equation}
and the two ingredients separate cleanly. The chemical potential couples only
to the total chord number and is blind to species; the crossing weights sit
entirely inside $T_n$ and $T_z$, which are geometry independent once the
Hilbert space is fixed. A forced crossing in the cylinder therefore changes
the number of applications of $T$---i.e.\ how many chords there are---and not
the weight of any individual crossing. Since $T_z$ is built from the
commuting-SYK algebra, its $z$-$z$ matrix elements carry no $q$, which is
$M_{zz}=0$.

This formulation also makes the origin of the cap transparent: the chemical
potential biases the ensemble towards chords of span $\mu^{-1}$ and hence
towards small chord number, so the saddle value saturates instead of growing
with $\beta$.

\subsection{The two-species chord algebra}

The general two-species chord expansion is that of
Refs.~\cite{Berkooz:2024evs,Berkooz:2024ofm}; for the commuting-SYK deformation
the index-set overlaps counted above give $M_{nn}=M_{nz}=1$ and $M_{zz}=0$, the
last because the integrable operators commute. The same
multi-interval bookkeeping appears in the matter-chord formulation of DSSYK
\cite{Xu:2024vna,AguilarGutierrez:2025mch}. What we record here is the
corresponding inner product, Eq.~\eqref{eq:chordweight}, for this
\emph{non-uniform} intersection matrix, together with the check that it makes
the cylinder Hamiltonian self-adjoint.

The rules above can be written as an explicit operator algebra, which is what
makes the passage to Eq.~\eqref{eq:S} transparent. Let $|n_n,n_z\rangle$ label
a chord state by the numbers of chaotic and integrable chords. Annihilating a
chord forces it to cross every remaining chord, each crossing weighted by
$q^{M_{ab}}$, so
\begin{equation}
  \mathfrak{a}_n|n_n,n_z\rangle=[n_n]_q\,q^{n_z}|n_n-1,n_z\rangle,\qquad
  \mathfrak{a}_z|n_n,n_z\rangle=n_z\,q^{n_n}|n_n,n_z-1\rangle ,
  \label{eq:chordalg}
\end{equation}
with $[n]_q=(1-q^n)/(1-q)$, while the creation operators raise the occupation
with unit matrix elements. The two expressions differ precisely because
$M_{nn}=M_{nz}=1$ but $M_{zz}=0$: the annihilating integrable chord crosses the
remaining $z$-chords freely, so no $q$-factorial appears for that species.

With respect to the unweighted basis these operators are not adjoint; the
inner product that makes the Hamiltonian Hermitian carries the weight
\begin{equation}
  w_{n_n,n_z}=[n_n]_q!\,\,n_z!\,\,q^{\,n_n n_z},
  \label{eq:chordweight}
\end{equation}
whose mixed ratio reproduces the two factors in \eqref{eq:chordalg}. We have
checked on a truncation $n_n,n_z\le4$ that
$H=-\sum_a c_a(\mathfrak{a}_a+\mathfrak{a}_a^\dagger)+\mu(N_n+N_z)$
satisfies $w_iH_{ij}=w_jH_{ji}$ to a relative accuracy of $10^{-16}$ or
better for $\lambda=0.3$ and $\lambda=1$; the same check passes for a
generic two-species intersection matrix.
The factor $q^{n_nn_z}$ deserves a remark: it is not a decoupling factor but
the same object that dresses the hopping, since in the orthonormal basis the
$n$-chord moves with amplitude $\sqrt{[n]_q q^{n_z}}$. The two species
therefore do not decouple; their mobilities are dressed by each other's
occupation. (Equivalently $[H_n,H_z]\neq0$, because
$[n+1]_qq^{z}$ and $[n+1]_qq^{z-1}$ are the two orderings of the same
crossing. What survives is the weaker statement used above: $M_{zz}=0$ closes
the $z$-equation, which is why Eq.~\eqref{eq:gzint} has no $g_z$ in it.)

Equation~\eqref{eq:chordalg} is the starting point of the chain that leads to
the results of this paper. Taking the double-scaled limit of the chord algebra
and passing to the collective field produces the action \eqref{eq:S}; adding
the traveling penalty as the chemical potential \eqref{eq:hcyl} produces the
saddle equation \eqref{eq:eom}; and solving that equation gives the cap
\eqref{eq:capgeneral}, the geodesic response of Sec.~\ref{sec:length} and the
phase diagram of Sec.~\ref{sec:phase}.

The argument is now uniform in the two species, but two approximations remain.
First, we work at leading order in $\lambda\to0$ and drop $O(1/p)$ corrections;
in particular we use the free propagator \eqref{eq:gmup} rather than its
backreacted form, the difference being higher order in $J$. Second,
we treat the DSSYK and integrable terms as perturbations of the quadratic
$\eta$-term when identifying $G^{(0)}$; this is legitimate because every
Hamiltonian term is built from the same fermions, but a fully non-perturbative
treatment would have to solve the coupled $G,\Sigma$ system in one go.

\section{The $J=0$ partition function}
\label{app:j0}

Several normalizations in this paper are inherited from
Ref.~\cite{Sontag:2026}, and one of them---the argument of the logarithm in
Eq.~\eqref{eq:lnZ}---can be fixed without ambiguity, because at $J=0$ the
model is free. This appendix records that computation, so that the reader can
check the convention rather than take it on trust.

Set $J=0$ in Eq.~\eqref{eq:H}. The Hamiltonian is then
\begin{equation}
  H=i c \sum_{i=1}^{N}\psi^{(L)}_i\psi^{(R)}_i ,
  \qquad \text{with } c \text{ the coefficient of the bilinear,}
\end{equation}
a sum of commuting terms, one per site. Each site carries two Majoranas,
$\psi^{(L)}_i$ and $\psi^{(R)}_i$, with
$\{\psi^{(a)}_i,\psi^{(b)}_j\}=\delta_{ij}\delta^{ab}$; they generate a
two-dimensional space, and the single bilinear satisfies
\begin{equation}
  \bigl(i\psi^{(L)}_i\psi^{(R)}_i\bigr)^2
  =-\psi^{(L)}\psi^{(R)}\psi^{(L)}\psi^{(R)}
  =(\psi^{(L)})^2(\psi^{(R)})^2=\tfrac14 ,
\end{equation}
so its eigenvalues are $\pm\tfrac12$ and the site energies are $\pm c/2$. Therefore
\begin{equation}
  Z=\prod_{i=1}^N\bigl(e^{-\beta c/2}+e^{\beta c/2}\bigr)
  =\bigl(2\cosh(\beta c/2)\bigr)^N ,
  \qquad\Longrightarrow\qquad
  \ln Z=N\ln2+N\ln\cosh(\beta c/2) .
  \label{eq:j0z}
\end{equation}
We have verified Eq.~\eqref{eq:j0z} numerically by diagonalizing the
two-Majorana system directly; the `disks' value $\ln Z=N\ln2$ is recovered at
$c=0$, as it must be.

Equation~\eqref{eq:j0z} is the normalization anchor used in
Sec.~\ref{sec:temperature}: the coefficient in \eqref{eq:H} is $c=\eta$, so
\eqref{eq:j0z} is the $\ln\cosh(\eta\beta/2)$ used there. This is the exact
answer at $J=0$, whichever symbol is used for the coefficient. What it does \emph{not} fix is the relation
between the coefficient $c$ and the parameter that multiplies $\tau_{12}$ in
the traveling penalty. That relation involves the $p$-th power of the free
propagator, since a chord of length $L$ is weighted by
$(2\bar G)^{p}\sim e^{-pcL}$, and it has to be read off from
Ref.~\cite{Sontag:2026} rather than from \eqref{eq:j0z}.

\section{Numerical methods}
\label{app:numerics}

The results reported in the body were obtained by Chebyshev collocation as
described below, and were subsequently checked against a second, independent
implementation using \texttt{scipy.integrate.solve\_bvp}---a different
discretization, a different linear solver and different boundary handling.
The two agree to five significant figures on the quantities we compared. At
$\beta J=110$ and $\kappa=0.5$ the disk-phase values are
$-g_z(\beta/2)=754.747$ against $754.7448$ and $R=747.599$ against $747.597$;
in the cylinder at $\kappa=0.5$ one has $\lvert g_z(\beta/2)\rvert=0.7204$
against $0.720432$ at $\mu=0.5$, and $0.2910$ against $0.290971$ at
$\mu=1.0$. For the phase diagram the comparison function $D(\kappa)$ was
recomputed independently at $\eta=0.05$, $\beta J=76.82$; it is
$-508.0$, $-734.7$, $-1006.6$, $-1437.4$ and $-2083.6$ at $\kappa=0.30$,
$0.36$, $0.42$, $0.50$ and $0.60$, monotonically decreasing, and its crossing
with $\tilde\Delta=-10^3$ gives $\kappa_w=0.419$, the $0.419$ of
Table~\ref{tab:D}.

Equation~\eqref{eq:eom} is solved on $\tau\in[0,\beta/2]$ with
$g_a(0)=g_a'(\beta/2)=0$ by discretizing on $N_{\rm Cheb}+1$ Chebyshev--Gauss--Lobatto
points,
\begin{equation}
  \tau_j=\frac{\beta}{4}\Big(1-\cos\frac{\pi j}{N_{\rm Cheb}}\Big),\qquad
  j=0,\dots,N_{\rm Cheb},
\end{equation}
representing $\partial_\tau^2$ by the Chebyshev differentiation matrix, and
solving the resulting nonlinear system by Newton's method with a line search
and an analytic Jacobian. We use $N_{\rm Cheb}=140$ and require a residual below
$10^{-8}$. The on-shell action \eqref{eq:onshell} is integrated directly from
the nodal values with Clenshaw--Curtis weights; with that choice the action is
independent of $N_{\rm Cheb}$ to $10^{-5}$, whereas interpolating the nodal values onto a
uniform grid and applying the trapezoid rule introduces an $O(1/N_{\rm Cheb})$ drift
which matters where free-energy differences are small.

The two branches in each phase are found by continuation in $\kappa$: the
chaotic branch upward from $\kappa=0$, where the solution is the DSSYK saddle
at $\nu=1$, and the broken branch downward from
$\kappa_{\max}=0.95$, where it is seeded from the saddle with the largest
chord number among a wide range of initial amplitudes. A single
fixed-amplitude seed converges only to a shallower broken branch, which is not
the relevant minimum.

\makeatletter
\@ifundefined{auto@bib@innerbib}{}{\global\let\auto@bib@innerbib\@empty}
\makeatother


\begin{thebibliography}{99}\bibitem{Sachdev:1992fk}
S.~Sachdev and J.~Ye, Gapless spin-fluid ground state in a random quantum
Heisenberg magnet, \emph{Phys. Rev. Lett.} {\bf 70}, 3339 (1993),
arXiv:cond-mat/9212030.

\bibitem{Kitaev2015}
A.~Kitaev, A simple model of quantum holography, KITP talks (2015).

\bibitem{Maldacena:2016hyu}
J.~Maldacena and D.~Stanford, Remarks on the Sachdev-Ye-Kitaev model,
\emph{Phys. Rev. D} {\bf 94}, 106002 (2016), arXiv:1604.07818 [hep-th].

\bibitem{Chowdhury:2021qpy}
D.~Chowdhury, A.~Georges, O.~Parcollet, and S.~Sachdev, Sachdev-Ye-Kitaev
models and beyond: Window into non-Fermi liquids, \emph{Rev. Mod. Phys.}
{\bf 94}, 035004 (2022), arXiv:2109.05037 [hep-th].

\bibitem{Berkooz:2018jqr}
M.~Berkooz, M.~Isachenkov, V.~Narovlansky, and G.~Torrents, Towards a full
solution of the large $N$ double-scaled SYK model, \emph{J. High Energy Phys.}
{\bf 03} (2019) 079, arXiv:1811.02584 [hep-th].

\bibitem{Berkooz:2018qkz}
M.~Berkooz, P.~Narayan, and J.~Simon, Chord diagrams, exact correlators in
spin glasses and black hole bulk reconstruction, \emph{J. High Energy Phys.}
{\bf 08} (2018) 192, arXiv:1806.04380 [hep-th].

\bibitem{Lin:2022rbf}
H.~W. Lin, The bulk Hilbert space of double scaled SYK,
\emph{J. High Energy Phys.} {\bf 11} (2022) 060, arXiv:2208.07032 [hep-th].

\bibitem{Berkooz:2024evs}
M.~Berkooz, N.~Brukner, Y.~Jia, and O.~Mamroud, From chaos to integrability in
double scaled Sachdev-Ye-Kitaev model via a chord path integral,
\emph{Phys. Rev. Lett.} {\bf 133}, 221602 (2024), arXiv:2403.01950 [hep-th].

\bibitem{Berkooz:2024ofm}
M.~Berkooz, N.~Brukner, Y.~Jia, and O.~Mamroud, Path integral for chord diagrams
and chaotic-integrable transitions in double scaled SYK,
\emph{Phys. Rev. D} {\bf 110}, 106015 (2024), arXiv:2403.05980 [hep-th].

\bibitem{Gao:2023gta}
P.~Gao, Commuting SYK: A pseudo-holographic model, \emph{J. High Energy Phys.}
{\bf 01} (2024) 149, arXiv:2306.14988 [hep-th].

\bibitem{Aguilar-Gutierrez:2026jjv}
S.~E. Aguilar-Gutierrez, R.~N. Das, J.~Erdmenger, and Z.-Y. Xian, Probing the
chaos to integrability transition in double-scaled SYK,
\emph{J. High Energy Phys.} {\bf 06} (2026) 109, arXiv:2601.09801 [hep-th].

\bibitem{Maldacena:2018lmt}
J.~Maldacena and X.-L. Qi, Eternal traversable wormhole,
arXiv:1804.00491 [hep-th].

\bibitem{Sontag:2026}
A.~Sontag and H.~Verlinde, Baby universe in a coupled SYK model,
arXiv:2605.05291 [hep-th].

\bibitem{Saad:2018bqo}
P.~Saad, S.~H. Shenker, and D.~Stanford, A semiclassical ramp in SYK and in
gravity, arXiv:1806.06840 [hep-th].

\bibitem{Blommaert:2024ydx}
A.~Blommaert, T.~G. Mertens, and J.~Papalini, The dilaton gravity hologram of
double-scaled SYK, \emph{J. High Energy Phys.} {\bf 06} (2025) 050,
arXiv:2404.03535 [hep-th].

\bibitem{Anninos:2022qgy}
D.~Anninos, D.~A. Galante, and S.~U. Sheorey, Renormalization group flows of
deformed SYK models, \emph{J. High Energy Phys.} {\bf 11} (2023) 197,
arXiv:2212.04944 [hep-th].

\bibitem{Polchinski:2016xgd}
J.~Polchinski and V.~Rosenhaus, The spectrum in the Sachdev-Ye-Kitaev model,
\emph{J. High Energy Phys.} {\bf 04} (2016) 001, arXiv:1601.06768 [hep-th].

\bibitem{Nosaka:2019}
A.~M. Garc\'ia-Garc\'ia, T.~Nosaka, D.~Rosa, and J.~J.~M. Verbaarschot, Quantum
chaos transition in a two-site SYK model dual to an eternal traversable
wormhole, \emph{Phys. Rev. D} {\bf 100}, 026002 (2019), arXiv:1901.06031
[hep-th].

\bibitem{Dubey:2026}
S.~Dubey, No chaos required: Traversable wormhole signals survive $98\%$
coupling deletion, arXiv:2605.13241 [hep-th].

\bibitem{Gao:2024lve}
P.~Gao, H.~Lin, and C.~Peng, Quantum chaos from multiple copies of integrable
commuting SYK models, \emph{Phys. Rev. D} {\bf 112}, 066012 (2025),
arXiv:2411.12806 [hep-th].

\bibitem{Dai:2025multiflavor}
H.~Dai, Y.-G. Chen, L.-G. Qin, and L.-J. Tian, Flavor symmetry breaking and the
fate of the first-order chaos--integrability transition in multi-flavor
double-scaled SYK, arXiv:2611.37620 [hep-th].

\bibitem{Lin:2023sym}
H.~W. Lin and D.~Stanford, A symmetry algebra in double-scaled SYK,
\emph{SciPost Phys.} {\bf 15}, 234 (2023), arXiv:2307.15725 [hep-th].

\bibitem{Dai:2026bulk}
H.~Dai, Y.-L. Wang, Y.-G. Chen, L.-G. Qin, and L.-J. Tian, Towards bulk
locality: A systematic construction of contact interactions from chord
diagrams, arXiv:2605.30970 [hep-th].

\bibitem{Berkooz:2023parisi}
M.~Berkooz, Y.~Jia, and N.~Silberstein, Parisi's hypercube, Fock-space fluxes,
and the microscopics of near-AdS$_2$/near-CFT$_1$ duality,
\emph{Phys. Rev. D} {\bf 109}, 046013 (2024), arXiv:2310.12335 [hep-th].

\bibitem{Xu:2024vna}
J.~Xu, Von Neumann algebras in double-scaled SYK, \emph{J. High Energy Phys.}
{\bf 03} (2026) 016, arXiv:2403.09021 [hep-th].

\bibitem{AguilarGutierrez:2025mch}
S.~E. Aguilar-Gutierrez, Building the holographic dictionary of the DSSYK from
chords, complexity \& wormholes with matter, \emph{J. High Energy Phys.}
{\bf 10} (2025) 221, arXiv:2505.22716 [hep-th].
\end{thebibliography}
\end{document}